\documentclass[a4paper,11pt]{article}
\usepackage[english]{babel}
\usepackage[utf8]{inputenc}
\usepackage{color}
\usepackage{amsmath,amsfonts}
\usepackage{amssymb}
\usepackage{subcaption}
\usepackage{graphicx}
\color{black}

\newtheorem{definition}{\bf Definition}[section]

\newtheorem{theorem}{\bf Theorem}[section]

\newtheorem{proposition}{\bf Proposition}[section]

\begin{document}

\title{A new integrable model of long wave-short wave interaction and linear stability spectra}

\author{
Marcos Caso-Huerta \\ Department of Mathematics, Physics and Electrical Engineering \\ Northumbria University, Newcastle upon Tyne, UK \\ Email: marcos.huerta@northumbria.ac.uk \\ {} \\
Antonio Degasperis \\ Dipartimento di Fisica \\ ``Sapienza'' Universit\`a di Roma, Rome, Italy \\ Email: antonio.degasperis@uniroma1.it \\ {} \\
Sara Lombardo \\ Department of Mathematical Sciences, School of Science \\ Loughborough University, Loughborough, UK \\ Email: s.lombardo@lboro.ac.uk \\ {} \\
Matteo Sommacal \\ Department of Mathematics, Physics and Electrical Engineering \\ Northumbria University, Newcastle upon Tyne, UK \\ Email: matteo.sommacal@northumbria.ac.uk
}

\date{}

\maketitle

\begin{abstract}
We consider the propagation of short waves which generate waves of much longer (infinite) wave-length. Model equations of such long wave-short wave resonant interaction, including integrable ones, are well-known and have received much attention because of their appearance in various physical contexts, particularly fluid dynamics and plasma physics. Here we introduce a new long wave-short wave integrable model which generalises those first proposed by Yajima-Oikawa and by Newell. By means of its associated Lax pair, we carry out the linear stability analysis of its continuous wave solutions by introducing the \emph{stability spectrum} as an  algebraic curve in the complex plane. This is done starting  from the construction of the eigenfunctions of the linearised long wave-short wave model equations. The geometrical features of this spectrum are related to the stability/instability properties of the solution under scrutiny. Stability spectra for the plane wave solutions are fully classified in the parameter space together with types of modulational instabilities.
\end{abstract}

\section{Integrable models of long wave-short wave resonant interaction}
\label{sec:into}
In several physical applications, waves are
represented as solutions of nonlinear partial differential equations. Finding out the dynamical behaviour of waves usually poses mathematical problems, which are very rarely solvable by analytical methods.
More frequently, approximate solutions may be found by treating nonlinear terms of the equations of motion as small perturbations of the linear part. In this framework, investigating how the linear superposition of  two or more  plane waves changes due to nonlinear effects leads to predictions on wave-wave interaction.
A well known approach of this kind, the multiscale method (see \textit{e.g} \cite{dodd1982}, \cite{degasperis2009} and the references quoted therein), requires that Fourier wave amplitudes be small, say $\mathcal{O}(\varepsilon)$, and be functions only of appropriately $\varepsilon$-rescaled space and time coordinates, where $\varepsilon$ is a small dimensionless parameter.
In this way, one obtains partial differential equations in the rescaled variables which are generally simpler, and sometimes even integrable.
When they are, these approximate wave equations, even if nonlinear, are exactly solvable by spectral techniques.

The best known example of such model equations is the nonlinear Schr\"{o}dinger (NLS) equation in a one-dimensional space,
\begin{equation}
\label{NLS}
iS_{t_2} + S_{\xi \xi} -2 \sigma S^{\ast}S^2 =0,\quad \sigma=\pm 1\;,
\end{equation}
where $\xi=\varepsilon (x-vt)$ and $t_2=\varepsilon^2 t$ are the rescaled space and time coordinates, and the asterisk indicates complex conjugation.
This equation follows via the multiscale method from almost any real propagation equation and provides the lowest order effect of the nonlinear terms on the linear solution
\[
\varepsilon S(\xi,t_2) e^{i(kx-\omega t)} + \varepsilon S^\ast(\xi,t_2) e^{-i(kx-\omega t)},\quad t_2=\varepsilon^2 t\,,\quad\xi=\varepsilon (x-vt),
\]
given by the sum of two plane waves with amplitudes $S$ and $S^\ast$, and with wave number $k$, frequency $\omega$ and group velocity $v$.

If more plane waves are superimposed, cross-interaction takes place if  the wave numbers and frequencies satisfy certain resonance conditions.
Among the many cases of interest and physical significance, we consider here wave equations which model
the interaction of two small-amplitude plane waves: one with a very long wave length ($k=0$) and a real amplitude $L$, and a second one with a much shorter wave length and a complex amplitude $S$.
Equations of this kind have been derived and proposed with various motivations, which mainly come from plasma physics \cite{yajima1976} and fluid dynamics \cite{benney1977,newell1978}.

On the mathematical side, long wave-short wave (LS) equations have been differently introduced because they are, at the same time, \emph{close} to physical equations and \emph{integrable}. This second distinctive property allows the construction of a broad range of solutions with relevant and useful insight into experimental observations. The focus of the present work is on LS models which are integrable.

 Two very well know examples of integrable LS models are the Yajima-Oikawa (YO) equation \cite{yajima1976}
\begin{equation}
\label{YO}
i S_t +S_{xx} - LS=0\,,\quad L_t=2\left(|S|^2\right)_x\,,
\end{equation}
and the alternative integrable Newell (N) wave system, introduced  in \cite{newell1978},
\begin{equation}
\label{N}
i S_t +S_{xx} +\left( i L_x + L^2 -2\sigma |S|^2\right) S=0\,,\quad
L_t=2\sigma \left(|S|^2\right)_x \,,\quad \sigma^2=1\,,
\end{equation}
where, in addition to a long wave-short wave  coupling, the short wave has the same self-interaction as the NLS equation, see (\ref{NLS}). The two systems (\ref{YO}) and (\ref{N}) are related by a Miura transformation, see \cite{ling2011}.

In \cite{calogero2000,calogero2001,degasperis2009} the YO equation has also been derived via multiscale method applied to a generic real wave equation, with the
resonance condition that the the group velocity $v_S=\omega'(k_S)$ of the short wave equals the group velocity $v_L=\omega'(0)$ of the long wave, say $v_S=v_L=v$, where $\omega(k)$ is the linear dispersion relation, $\omega'(k)=\mathrm{d}\omega(k)/\mathrm{d}k$, and $k_S$ is the non-vanishing wave number of the short wave.

The integrability of both the YO and N equations has been exploited to investigate extensions in various directions.
Thus these LS wave equations have been generalised to multi-component short waves, either in vector form, as in \cite{geng2020}, or in matrix form as in \cite{geng2020a} requiring a higher rank matrix Lax pair. A third LS wave scalar equation, which reads
\begin{equation}
\label{LS2}
i S_t +S_{xx} + i\left(LS\right) _x -2 |S|^2 S=0\,,
\quad L_t=2 \left(|S|^2\right)_x  \,,
\end{equation}
has been derived in \cite{geng2019} from a $3\times 3$ matrix Lax pair, which is however different from the Lax pair investigated in the present  paper (see Sections \ref{sec:integrable} and \ref{sec:interaction}), and is not discussed here.

In Section \ref{sec:interaction}, we show that the YO and N equations, (\ref{YO})  and (\ref{N}), do not need to be treated separately. Indeed, these two model equations can be combined in just one system, namely
\begin{equation}
\label{YON}
i S_t +S_{xx} +\left(i\alpha L_x+\alpha^2 L^2-\beta L -2\alpha |S|^2 \right) S=0\,,
\quad L_t=2 \left(|S|^2\right)_x \,,
\end{equation}
which we refer to as YON equations. This system coincides with the YO equation (\ref{YO}) for $\alpha=0$, $\beta=1$, while it reads as the N equation (\ref{N}) by setting $\alpha=\sigma$, $\beta=0$ and by substituting the field $L$ with $\sigma L$ ($\sigma=\pm 1$).
In fact, as we will show in Section \ref{sec:interaction}, the novel system (\ref{YON}) is integrable for \emph{any} real value of $\alpha$ and $\beta$. Thus it is likely to be relevant to applications as it is more flexible in modelling the long wave-short wave interaction.
Moreover, this unifying result makes our present overall analysis and discussion of the two model equations (\ref{YO}) and (\ref{N}) simpler and more compact.

Other LS wave integrable equations may be sorted out via transformations of the wave fields. Indeed, the two integrable LS equations (\ref{YON}) and (\ref{LS2}) may be given an equivalent, even simpler, form by performing the gauge transformation
\cite{newell1978,geng2019}
\begin{equation}
\label{gauge}
S(x,t)= e^{i\phi(x,t)} \hat{S}(x,t)\;,\quad L(x,t)=\hat{L}(x,t),\quad \phi_x=\mu L,\quad \phi_t=2\mu\left(|S|^2\right)\,.
\end{equation}
This transformation, which introduces the extra real parameter $\mu$, originates from the conservative form of the evolution equation for the long wave amplitude $L$, in both the systems (\ref{YON}) and (\ref{LS2}). Thus gauge transforming the YON system (\ref{YON})
yields the three parameter family of LS wave resonance equations
\begin{equation}
\label{YONgauge}
i \hat{S}_t +\hat{S}_{xx} + 2i\mu \hat{L} \hat{S}_x + \left\{ \left(\alpha+\mu\right) \left[ i \hat{L}_x+
\left(\alpha-\mu \right) \hat{L}^2 -2 |\hat{S}|^2 \right ] -\beta \hat{L}\right\}\hat{S}=0\,,
\quad \hat{L}_t=2 (|\hat{S}|^2)_x\,,
\end{equation}
while the same transformation, as applied to the integrable equations (\ref{LS2}), leads to the LS wave system
\begin{equation}
\label{LS2gauge}
i \hat{S}_t +\hat{S}_{xx} + i\mu \hat{L} \hat{S}_x +\left(\mu +1\right) \left[i \left(\hat{L} \hat{S}\right)_x- \left (\mu \hat{L}^2 + 2 |\hat{S}|^2 \right) \hat{S}\right ]=0\,,\quad
\hat{L}_t=2 (|\hat{S}|^2)_x\,.
\end{equation}
The integrability of (\ref{YO}), (\ref{N}) and (\ref {LS2}), and of their transformed versions via (\ref {gauge}), allows the construction of special exact solutions of these model equations, such as solitons and rogue waves
(\textit{e.g.}, see \cite{ma1978,baronio2015,chen2018,chen2019,geng2020}).

Integrability is also a key property to address one of the basic issues in nonlinear wave dynamics, namely that of linear stability against small changes of the initial conditions (\textit{e.g.}, see \cite{degasperis2018} and the references quoted there).
For instance, the modulational instability of a periodic wave train, as in fluid dynamics and optics, has been well exemplified via the NLS equation (\ref{NLS}) with focusing self-interaction ($\sigma=-1$). Indeed, the NLS equation has deserved special interest as it proves to yield, for focusing interaction, a simple description of the instability of a regular wave train, a phenomenon first observed by Talanov \cite{talanov1965} in optics and by Benjamin-Feir \cite{benjamin1967} in a water tank. For this particular equation, understanding modulational instability of continuous wave solutions may be achieved via standard Fourier analysis. However this analysis fails if the unperturbed solution is not just a plane wave (see \cite{pelinovsky2021}), or if several nonlinear wave trains have a resonant interaction.
In general, when dealing with integrable systems, the stability of interacting plane waves is better treated by means of integrability techniques rather than by a Fourier approach. Indeed, the Lax pair plays a key-role as it allows to extend the linear stability analysis to other wave solutions,  such as those constructed from plane waves by Darboux dressing methods \cite{wright2010}.

The integrability method  has been used to unveil the stability properties of plane wave solutions of two coupled NLS equations \cite{degasperis2018,degasperis2019}. For these model equations, instabilities have been fully classified in terms of coupling constants, amplitudes and wave numbers, including instability effects due to defocusing self- and cross-interactions. As for the long wave-short wave resonance interaction, in addition to orbital stability \cite{borluk2008} and transverse stability \cite{erbay2012}, also linear stability has been considered for solutions of particular models \cite{newell1978}.

On the wake of those results, we investigate here the linear stability of continuous wave solutions of the YON equations (\ref{YON}). Although  the presentation is intended to be self-contained, we refer the reader to \cite{degasperis2018,degasperis2019} for specific proofs and further details on the general method. The main target of our approach is the computation of the \emph{stability spectrum} ($\mathbb{S}$), associated to the given plane wave solution.
 For a multicomponent system such as (\ref{YON}), this spectrum does not generically coincide with the ordinary Lax spectrum.
It turns out to be a piece-wise continuous curve in the complex plane of the spectral variable $\lambda$.
In each point of this curve, one or more eigenmode solutions of the linearised equation of motion are well defined, see Section \ref{sec:integrable}.
The relevant stability-instability properties of the wave solution are readily conveyed by the geometrical properties of the spectral curve $\mathbb{S}$ itself. Additional relevant targets are the linearised equation eigenfunction frequencies and their associated  \emph{gain} functions, which tell us whether the wave number instability bands are of \emph{passband} or \emph{baseband} type \cite{baronio2014}.

This paper is organised as follows. In Section \ref{sec:integrable}, we review our approach to linear stability; in particular, we sketch the role that integrability plays in investigating the linear stability of a given solution. In Section \ref{sec:interaction}, the integrability of the unified YON model is established by showing its corresponding Lax pair. Moreover, we consider the plane wave solution of the YON model equations, and study its linear stability by constructing and classifying the corresponding stability spectra.
We compute the gain function, characterise the instabilities and show that the plane wave solution is unstable for a generic choice of the physical parameters. We conclude with some remarks and outlooks in Section \ref{sec:conclusions}.

 \section{Integrable equations and small variation of their solutions}
 \label{sec:integrable}
Our approach to the linear stability of solutions of integrable partial differential equations starts from the associated Lax pair, a characteristic feature of integrable systems:
\begin{equation}\label{laxpair}
\Psi_x=X\,\Psi\,, \quad \Psi_t=T\,\Psi\,.
\end{equation}
In the present context, the solution $\Psi(x,t,\lambda)$ and the coefficients $X(x,t,\lambda)$, $T(x,t,\lambda)$ are $3\times3$ matrices.
The latter two matrices are assumed to have the following  polynomial dependence on the complex \emph{spectral variable} $\lambda$:
\begin{equation}
\label{XTpair}
X(\lambda)=i \lambda \Sigma + Q\,,\quad T(\lambda)= (i \lambda)^2 A + i\lambda B + C\,,
\end{equation}
where $\Sigma$ is the constant, traceless, diagonal matrix
\begin{equation}
\label{matrixsigma}
\Sigma=\textrm{diag}\{1 \,,\, 0\,,\, -1 \}=\left(\begin{array}{ccc} 1 & 0 & 0 \\  0 & 0 &  0  \\ 0 & 0  & -1 \end{array} \right),
\end{equation}
and $Q(x,t)$ is $\lambda$-independent and off-diagonal, namely $Q_{jj}=0$, while its off-diagonal entries $Q_{jm}(x,t)$, $j\neq m$, are generically six complex-valued functions of $x$ and $t$;  $A$ is a constant matrix, while $B$ and $C$ can be written in terms of commutators and anti-commutators of $\Sigma$, $Q(x,t)$ and their powers.
Then, the compatibility condition
\begin{equation}
\label{compat}
X_t - T_x + [X\,,\,T] =0
\end{equation}
leads to an evolution equation for the matrix $Q$ in the form of a system of six coupled nonlinear wave equations for the entries of $Q$.
Among other reductions, this system can be reduced to just two equations describing the resonant interaction of two waves of physical interest (see next section).

Before looking at any particular reduction, let us first turn our attention to the way integrability plays an essential role in investigating the linear stability of a given solution $Q(x,t)$. In order to make this paper self-contained, we first briefly report the main ingredients and features of the method (see \cite{degasperis2018} and references therein).
The starting point is the following theorem (see \cite{degasperis2018} for a proof), which provides the quite remarkable and well known property of the matrix
solution $\Psi(x,t,\lambda)$ of the two ordinary differential equations (\ref{laxpair}):
\begin{theorem}
\label{th:1}
Let $Q(x,t)$ be a given solution of the compatibility equation (\ref{compat}) and let $\Psi(x,t,\lambda)$ be a corresponding fundamental solution of (\ref{laxpair}), that is $\det\Psi \neq 0$, then the matrix
 \begin{equation}
 \label{Fmatrix}
F(x,t,\lambda)= \left[ \Sigma\,,\,\Psi(x,t,\lambda) \,M(\lambda) \,\Psi^{-1}(x,t,\lambda) \right]\,,
\end{equation}
where $M(\lambda)$ is an arbitrary $(x,\,t)$-independent matrix, is a solution of the \emph{linear} equation obtained by linearising around $Q(x,t)$ the nonlinear evolution equation resulting from (\ref{compat}).
 \end{theorem}
In other words, suppose that $Q(x,t)$ is such a given solution, and that $Q(x,t) + \delta Q(x,t)$ is a second solution of the same evolution equation. Suppose moreover that the entries of $\delta Q(x,t)$ are sufficiently small (this is certainly so at small initial times $t$, if the initial condition $\delta Q(x,0)$ is small enough), then the linearised evolution equation for $\delta Q(x,t)$ is obtained by neglecting all the nonlinear terms.
This evolution equation  takes the form
\begin{equation}
\label{linear}
\delta Q_t(x,t)= \mathcal{L} (\delta Q(x,t))\,,
\end{equation}
where $\mathcal{L}(V)$ is a matrix-valued linear function (with $(x,t)$-dependent coefficients) of $V$ and its $x$-derivatives.
Then Theorem \ref{th:1} states that the matrix $F(x,t,\lambda)$ (\ref{Fmatrix}) is a $\lambda$-dependent solution of the same linear equation (\ref{linear}), namely
\begin{equation}
\label{Flinear}
F_t(x,t,\lambda)= \mathcal{L} (F(x,t,\lambda))\,,
\end{equation}
for any complex $\lambda$.

This is a system of six linear partial differential equations; its solutions $F(x,t,\lambda)$, which depend on the spectral variable $\lambda$, play the role of Fourier-like modes if the values of $\lambda$ lie in an appropriate subset $\mathbb{S}$ of the complex plane, defined below as \emph{stability spectrum} because of its key relevance to finding out whether a given solution $Q(x,t)$ is linearly stable or unstable.
This spectrum, which depends on the particular solution $Q(x,t)$, is defined so as to guarantee the boundedness of each solution $F(x,t,\lambda)$, at any fixed $t$, on the entire $x$-axis.
By keeping in mind this analogy with the Fourier analysis of solutions of linear partial differential equations with constant coefficients, we assume that the set of solutions $F(x,t,\lambda)$, see (\ref{Fmatrix}), of the equation (\ref{Flinear}) allows the representation
\begin{equation}
\label{fourier}
\delta Q(x,t)= \int_{\mathbb{S}} \mathrm{d}\lambda\, F(x,t,\lambda)\,,
\end{equation}
for an appropriate choice of $M(\lambda)$, and in an appropriate functional space of initial data: for instance, we have in mind the physical assumption that the entries of the initial condition $\delta Q(x,0)$ be localised wave packets.
Although the proof \cite{degasperis2018} of the theorem above requires only local properties (differentiability up to sufficiently high order in $x$), the integral representation (\ref{fourier}) should be justified by spectral methods.
To our present purpose, that is assessing the linear stability of $Q(x,t)$, we do not need to enter into this matter, while we devote our next section to detail the properties of the stability spectrum $\mathbb{S}$, for a particularly simple choice of $Q(x,t)$.

Before going into this investigation, we give an explicit and general construction of the matrix eigenfunctions $F(x,t,\lambda)$ by taking advantage of the properties of the wedge product of 3-dimensional vectors.
Consider first the matrix solution $\Psi$ of the Lax pair (\ref{laxpair}) and its three column vectors $\psi^{(j)}$, $j=1, 2, 3$:
\begin{equation}
\label{matrixPSI}
\Psi=\left(\begin{array}{ccc}\psi^{(1)} & \psi^{(2)} &  \psi^{(3)} \end{array} \right )\;\;.
\end{equation}
Without loss of generality, we assume that this fundamental solution has unit determinant,
say $\det\,\Psi= \psi^{(1)}\cdot \psi^{(2)}\wedge \psi^{(3)}=1$.
This implies that we can give the defining expression (\ref{Fmatrix}) of the matrix eigenmode the alternative expression
\begin{equation}
\label{FFmatrix}
F(x,t,\lambda)= \left[ \Sigma\,,\,\Psi(x,t,\lambda) \,M(\lambda)\, \Psi^{A}(x,t,\lambda) \right]\,,
\end{equation}
where $\Psi^{A}$ is the adjugate matrix (or classical adjoint) of $\Psi$. In turn, the matrix of co-factors $\Psi^{A}$
can be expressed through its rows, namely
\begin{equation}\label{matrixPSIA}
\Psi^{A}=\left (\begin{array}{ccc}  {\psi^{A(1)}}^{T} \\  {\psi^{A(2)}}^{T} \\  {\psi^{A(3)}}^{T} \end{array} \right )\;\; ,
\end{equation}
where the superscript $T$ stands for transposition that takes a column vector $\psi^{A(j)}$ into a row vector, so that the vectors $\psi^{A(j)}$ are the columns of ${\Psi^{A}}^{T}$.
The computation of $\Psi^{A}$ is then provided by the expression (\ref{matrixPSIA}) with
\begin{equation}
\label{rowPSIA}
\psi^{A(j)}= \psi^{(m)} \wedge \psi^{(n)}\,,\quad \{j, m, n\}= \text{cyclic permutations of} \;\{1, 2, 3\}\,.
\end{equation}
As we need six eigenfunctions for each value of $\lambda$ if $Q(x,t)$ is generic (non-reduced), we choose to compute the eigenfunctions (\ref{FFmatrix}) by setting $M$ equal to  one of the six  basis matrices $M^{(jm)}$,
\begin{equation}
\label{Mbasis}
M^{(jm)}_{ab}= \delta_{ja} \delta_{mb}\,,\quad j,m=1,2,3\,,\quad j\neq m\,,
\end{equation}
of the space of the off-diagonal $3\times 3$ matrices (here $\delta_{jm}=1$ for $j=m$ and $\delta_{jm}=0$ for $j\neq m$), namely $F^{(jm)}=\left[ \Sigma\,,\,\Psi M^{(jm)} \Psi^{A} \right]$.
The convenience of this choice stems from the algebraic identity
\begin{equation}
\label{eigenF}
\Psi(x,t,\lambda)\,M^{(jm)}\, \Psi^{A}(x,t,\lambda)  = \psi^{(j)} \left( \psi^{(j)} \wedge \psi^{(n)} \right)^T \;\epsilon_{jnm}\;\;,
\end{equation}
where $\epsilon_{jnm}$
is the parity of the permutation $\{j, n, m\}$ of $\{1, 2, 3\}$, that is $\epsilon_{jnm}=1$ if $\{j, n, m\}$ is a cyclic permutation of $\{1, 2, 3\}$, and
$\epsilon_{jnm}=-1$ otherwise.
Thus, the six eigenfunctions finally take the expression
\begin{equation}
\label{eigenFF}
F^{(jm)}(x,t,\lambda)= \left[ \Sigma\,,\,\psi^{(j)} \left( \psi^{(j)} \wedge \psi^{(n)} \right)^T \right]\;\epsilon_{jnm}\;\;,\quad j\neq m\,,
\end{equation}
with the side remark that this explicit expression of the eigenfunction $F^{(jm)}$ cubically depends on just two vector solutions of the Lax pair. Finally, the Fourier-like representation (\ref{fourier}) can be more precisely written as
\begin{equation}
\label{fourierlike}
\delta Q(x,t)= \int_{\mathbb{S}} \mathrm{d}\lambda \sum_{j,m}\mu^{(jm)}(\lambda)F^{(jm)}(x,t,\lambda)\,,
\end{equation}
where the six functions $\mu^{(jm)}(\lambda)$ play the role of Fourier-like transform of $\delta Q$.

\section{Interaction of long and short continuous waves}
\label{sec:interaction}
In this section we consider reductions of the matrix $Q$ in (\ref{XTpair}) which lead, via the compatibility condition (\ref{compat}), to a system of two evolution equations able to model the long wave-short wave resonant interaction.
To this purpose, the
matrix $Q$ should depend only on two wave fields: the amplitudes $L(x,t)$, modulating a continuous wave with a (infinitely) long wave length, and $S(x,t)$, which modulates the amplitude of a periodic short wave.
As a consequence, one should impose the condition that $L(x,t)$ be real and dispersion-free, and that $S(x,t)$ be instead complex with a non-vanishing dispersion coefficient (which, without any loss of generality, is rescaled to unity).
As we pointed out in Section \ref{sec:into}, here we introduce the new integrable model (\ref{YON}), which includes the YO and N systems (\ref{YO}) and, respectively (\ref{N}), by setting
\begin{equation}
\label{Q}
Q=\left (\begin{array}{ccc} 0 & S & iL \\ \alpha  S^\ast & 0 &   S^\ast  \\ i\alpha^2L-i\beta & \alpha S & 0 \end{array} \right )\,.
\end{equation}
This system is integrable for \emph{any} real value of $\alpha$ and $\beta$; it reduces to the YO equation (\ref{YO}) for $\alpha=0$, $\beta=1$ and to the N equation (\ref{N}) for $\alpha=\sigma$, $\beta=0$ and $L\mapsto \sigma L$ ($\sigma=\pm 1$).

From this choice of the matrix $Q$, and from the compatibility condition (\ref{compat}), the coefficients $A$, $B$, $C$ of the matrix $T$ (see (\ref{XTpair})) turn out to be:
\begin{subequations}
\label{ABC}
\begin{equation}
\label{AB}
A=\frac{i}{3}\left (\begin{array}{ccc} 1 & 0 & 0 \\ 0 & -2 & 0  \\0 & 0 & 1 \end{array} \right )\,,
\quad B= \left (\begin{array}{ccc} 0 & i S & 0 \\ i \alpha S^\ast & 0 & -i S^\ast  \\ 0 & -i\alpha S & 0 \end{array} \right )\,,
\end{equation}
\begin{equation}
\label{C}
C= \left (\begin{array}{ccc} -i\alpha |S|^2 & -\alpha LS +iS_x & i|S|^2 \\ - \alpha^2 LS^\ast +\beta  S^\ast-i \alpha S^\ast_x  & 2i \alpha |S|^2 &  -\alpha LS^\ast -i S^\ast_x \\ i \alpha^2 |S|^2& -\alpha^2 LS +\beta S+i\alpha S_x & -i\alpha |S|^2 \end{array} \right )\,.
\end{equation}
\end{subequations}
In these formulae, the parameters $\alpha$, $\beta$ are arbitrary but  constrained to be real. These parameters may be considered as independent constants which are responsible for the long-short wave cross-interaction.

Here we are concerned with the linear stability of a given solution $S(x,t)$, $L(x,t)$ of the system (\ref{YON}), under the assumption that this solution is bounded for all values of $x$. According to standard practice, we proceed by investigating the time evolution of a small variation $\delta Q(x,t)$, namely $\delta S(x,t)$, $\delta L(x,t)$, of this solution.
At the early stage of the evolution these variations are assumed to be sufficiently small so as to keep only their linear contributions. Thus, they satisfy the linearised approximate equation (\ref{linear})
\begin{align}
\label{linearYON}
\begin{array}{l}
i\delta S_t +\!\delta S_{xx} +\!\left(i\alpha L_x+\alpha^2 L^2-\beta L -4\alpha |S|^2\right)\!\delta S +i\alpha S \delta L_x +\!\left(2\alpha^2 L  -\beta \right)\!S\delta L -2\alpha  S^2 \delta S^\ast\!\!=0 \\
\delta L_t=2 \left(S\delta S^\ast +S^\ast \delta S\right)_x \,.
\end{array}
\end{align}
Moreover, and for the sake of simplicity, we assume that the initial values $\delta S(x,0)$ and $\delta L(x,0)$ are localised and bounded functions of $x$.
With these assumptions in mind, the solution $S(x,t)$, $L(x,t)$ is linearly stable if $\delta S(x,t)$, $\delta L(x,t)$ remain small for all later times $t > 0$ or, equivalently because of the integral representation (\ref{fourierlike}), if the eigenfunctions $F^{(jm)}(x,t,\lambda)$, see (\ref{eigenFF}), do not grow exponentially in time.

As testing a recipe goes by tasting the cake, we apply our approach to the simplest physically relevant solution of the long wave-short wave equation (\ref{YON}), namely the continuous wave solution
\begin{equation}
\label{CW}
S(x,t)= a e^{i\theta},\quad L(x,t)  = b,\quad \theta = qx - \nu t,\quad \nu = q^2 - \alpha^2 b^2 + \beta b +2\alpha a^2 \,,
\end{equation}
which introduces three independent real parameters: the two amplitudes $a$ and $b$, and the wave number $q$ of the short wave.
The starting point is the construction of a corresponding fundamental solution $\hat{\Psi}(x,t,\lambda)$ of the Lax equations (\ref{laxpair}). This is given by the expression
\begin{equation}
\label{hatlaxsolution}
\hat{\Psi}(x,t,\lambda) = e^{i\rho(\lambda) t}R(x,t) e^{i (x W(\lambda) - t W^2(\lambda))},\quad R(x,t) = \text{diag}\{1,e^{-i\theta},1\}\,,
\end{equation}
where
\begin{equation}\label{rho}
\rho(\lambda)=\frac23 \lambda^2+\alpha^2 b^2-2\alpha a^2-\beta b,
\end{equation}
and where the $(x,t)$-independent matrix $W(\lambda)$ is
\begin{equation}
\label{Wmatrix}
    W(\lambda) = \begin{pmatrix} \lambda & -i a & b \\
    -i \alpha a & q & -i a \\
    \alpha^2 b- \beta & -i \alpha a & -\lambda
    \end{pmatrix}\,.
\end{equation}
Because of the trace expressions
\begin{equation}\label{traces}
\text{tr}(W)= q\,,\quad\text{tr}(W^2)= \nu +3 \rho\,,
\end{equation}
the matrix solution (\ref{hatlaxsolution}) has unit determinant, det$(\hat{\Psi}(x,t,\lambda)) = 1$. To our purpose, and according to the formalism detailed in the previous section, it is however more convenient to choose the alternative solution $\Psi(x,t,\lambda)$ whose column vectors, see (\ref{matrixPSI}), are
\begin{equation}
\label{columns}
\psi^{(j)}=\hat{\Psi}(x,t,\lambda) f^{(j)}(\lambda)\,,\quad j=1,\,2,\,3\,.
\end{equation}
Here the three constant vectors $f^{(j)}(\lambda)$ are the eigenvectors of the matrix $W(\lambda)$ (\ref{Wmatrix}),
\begin{equation}
\label{eigenvec}
W f^{(j)} = w_j f^{(j)} \,,\quad j=1,\,2,\,3\,,
\end{equation}
which are generically linearly independent. Moreover, we normalise them so that
\begin{equation}
\label{vectors}
 f^{(1)} \cdot \left( f^{(2)} \wedge f^{(3)} \right) =1\,,
\end{equation}
with the implication that det$(\Psi(x,t,\lambda)) = 1$.
This construction finally yields the expression of the column vectors of  $\Psi(x,t,\lambda)$, see (\ref{columns}), namely
\begin{equation}
\label{COLumns}
\psi^{(j)}=e^{i(\eta_j +\rho t)} R f^{(j)}\,,
\end{equation}
with
\begin{equation}
\label{phases}
 \eta_j =w_j x-w_j^2 t \,,\quad j=1\,,2\,,3\,.
\end{equation}
We are now in the position to obtain the expression of the eigenfunctions $F^{(jm)}(x,t,\lambda)$, which correspond to the continuous wave solution (\ref{CW}). By inserting in (\ref{eigenFF}) the
vectors (\ref{COLumns}), and by taking into account the relation
\begin{equation}
\label{etasum}
\eta_1+\eta_2+\eta_3= \theta -3\rho t\,,
\end{equation}
and the matrix identity (for two arbitrary vectors $u$, $v$)
\begin{equation}
\label{identity}
e^{ i \theta} (Ru) (Ru \wedge Rv)^T =R\,\left[u (u \wedge v)^T\right]\,R^{-1}\,,\quad R(x,t) = \text{diag}\left\{1,e^{-i\theta},1\right\}\,,
\end{equation}
we end up with the expression
\begin{equation}
\label{CWeigenFF}
F^{(jm)}(x,t,\lambda)= e^{ i(\eta_j-\eta_m)}\,
R\, \left[ \Sigma\,,\,f^{(j)} ( f^{(j)} \wedge f^{(n)} )^T \right]\, R^{-1} \,\epsilon_{jnm}\,,\quad j\neq m\,,
\end{equation}
where $\epsilon_{jnm}$ is the parity of the permutation $\{j,n,m\}$ of $\{1, 2, 3\}$.
This formula explicitly shows that, apart from the $\lambda$-independent phase $\theta$, the eigenfunctions $F^{(jm)}(x,t,\lambda)$ depend on $x$ and $t$ only via the exponentials $e^{ i(\eta_j-\eta_m)}$.
By taking into account the expression of $\eta_j$, see (\ref{phases}), these exponentials for $j \neq m$ take the familiar expression
\begin{equation}
\label{EXP}
 e^{\pm i(k_n x-\omega_n t)}\,,\quad k_n =w_{n+1}-w_{n+2}\,,\quad\omega_n =w^2_{n+1}-w^2_{n+2}\;,\quad n=1, 2, 3\;\; \text{mod}\, 3\;\;,
 \end{equation}
where the wave numbers $k_j(\lambda)$ and their corresponding frequencies $\omega_j(\lambda)$ are explicitly defined as
\begin{equation}
\label{k-omega}
\begin{array}{ccc} k_1=w_2-w_3 \;,& k_2=w_3-w_1 \;,& k_3=w_1-w_2 \;, \\
\\
\omega_1 = w^2_2-w^2_3 \;,& \omega_2=w^2_3-w^2_1 \;,& \omega_3=w^2_1-w^2_2 \,, \end {array}
\end{equation}
in terms of the eigenvalues $w_j$ of the matrix $W$, see (\ref{eigenvec}), or, equivalently, in terms of the roots of the characteristic polynomial
\begin{equation}
\label{Pw}
P(w, \lambda)= \det[w\, \mathbf{1}-W(\lambda)]=(w-w_1)(w-w_2)(w-w_3)= (w-q)(w^2-\lambda^2 + p)+r \;\;,
\end{equation}
where the two parameters $p$ and $r$ have the expression
\begin{equation}
\label{pr}
p=2 \alpha a^2 -\alpha^2 b^2+\beta b = \nu-q^{2}\,,\qquad  r= a^2 [2\alpha(q+\alpha b) - \beta]\,.
\end{equation}
Because of the requirement that the basic solutions (\ref{CWeigenFF}) be bounded functions of $x$,
we conclude that the spectral representation of $\delta L(x,t)$ and of $\delta S(x,t)$ requires integrating with respect to the complex spectral variable $\lambda$, see (\ref{fourierlike}), over the subset of the complex $\lambda$-plane where at least one of the wave numbers $k_j(\lambda)$ is real.
 These considerations then lead to the following
\begin{definition}
\label{def:spectrum}
The stability spectrum $\mathbb{S}$ is defined as the set of the complex values of $\lambda$ such that at least one of the three wave number functions $k_1(\lambda)$, $k_2(\lambda)$, $k_3(\lambda)$ is real.
\end{definition}
In this construction, we notice that the dispersion relation between the wave number $k_j$ and its corresponding frequency $\omega_j$ is \emph{parametrically} defined by the pair
of functions $k_j(\lambda)$ and $\omega_j(\lambda)$  by varying the parameter $\lambda$ only over the stability spectrum $\mathbb{S}$.
It follows that the continuous wave solution (\ref{CW}) is linearly stable if $\omega_j (\lambda)$ is real for those values of $j$ such that $k_j(\lambda)$ is real for any $\lambda$ over the entire spectrum $\mathbb{S}$.
On the contrary, if for some $j$ and $\lambda \in\mathbb{S}$, $\omega_j(\lambda)$ is not real,  then this solution is linearly unstable.
In this latter case, the relevant physical information is provided by the computation of the gain function
\begin{equation}
\label{gain}
\Gamma_j(\lambda) = |\text{Im}(\omega_j(\lambda))|\;\;,\quad \lambda \in \mathbb{S}\;\;,\quad\text{Im}(k_j(\lambda))=0\,\quad .
\end{equation}

Before proceeding to classify the stability spectra $\mathbb{S}$ and gain functions (\ref{gain}),
 it is worth drawing the reader's attention to the following fact. If the present approach to linear stability is applied to continuous wave solutions of the NLS equation (\ref{NLS}), then the stability spectrum introduced above by the definition (\ref{def:spectrum}) coincides with the Lax spectrum (which follows from the Lax differential operator in the space variable $x$ via standard definition) \cite{degasperis2016, pelinovsky2021}.
However, this coincidence of the two spectra is a peculiar property of the $2\times2$ matrix Lax equations.
In the case of $N\times N$ Lax equations with $N\geq 3$, the two spectra, i.e. the Lax spectrum and the stability spectrum, as defined here by (\ref{def:spectrum}) for $N=3$, are generically different from each other. For such larger matrices only the spectrum $\mathbb{S}$ as defined above is relevant to stability.

We begin our analysis of the stability spectra $\mathbb{S}$ and of the corresponding gains by observing that the expression of the coefficients of the characteristic polynomial $P(w, \lambda)$ (\ref{Pw}) depends on the wave number $q$ of the short wave solution (\ref{CW}), on the  parameter $r$ given in (\ref{pr}), and on the parameter $p$ (see (\ref{pr})) and the spectral variable $\lambda$ only via the combination $\lambda^2-p$.
This shows that it is sufficient to fix the value of the parameters $q$ and $r$ in the polynomial (\ref{Pw}) and to define the stability spectrum as a curve in the complex plane of the variable $\lambda^2-p$. This simplifies our task as the parameter space is reduced to the $(q, r)$ plane. Thus, hereafter, we introduce the alternative, and more convenient, complex variable $\Lambda$ defined as
\begin{equation}
\label{Lambda}
\Lambda= \lambda^2 -p\,,
\end{equation}
with the implication that the parameter $p$ becomes irrelevant to our characterisation and classification of spectra.
In fact, by a minor abuse of notation, we refer to the characteristic polynomial (\ref{Pw}) as
\begin{equation}
\label{PwLambda}
P(w, \Lambda)= \det[w\, \mathbf{1}-W]=(w-w_1)(w-w_2)(w-w_3)= (w-q)(w^2-\Lambda)+r \,.
\end{equation}
In order to make this change of variable from $\lambda$ to $\Lambda$ explicit, we denote as $\mathbb{S}^{\Lambda}$ the stability spectrum in the complex $\Lambda$-plane by adopting the following definition
\begin{definition}
\label{def:Spectrum}
The stability spectrum $\mathbb{S}^{\Lambda}$ is defined as the set of all complex values of $\Lambda$ such that at least one of the three wave number functions $k_1(\Lambda)$, $k_2(\Lambda)$, $k_3(\Lambda)$, see (\ref{k-omega}), is real.
\end{definition}
It is obvious that this spectrum looks different from the spectrum $\mathbb{S}$.
Furthermore, we note that the parameter $q$, if non-zero, can be rescaled to the value $q=1$ by rescaling $w$ by $q$, $\Lambda$ by $q^2$ and $r$ by $q^3$, or, equivalently, by the change of variables
\begin{equation}
\label{q1}
w\rightarrow qw\,,\quad \Lambda\rightarrow q^2 \Lambda\,,\quad r\rightarrow q^3r\,.
\end{equation}
Nevertheless, we find it convenient to keep $q$ in our formulae, to numerically set $q=1$ whenever $q\neq 0$,  and to separately treat the case $q=0$.

Let us consider now the part of the spectrum $\mathbb{S}^{\Lambda}$ which lies on the \emph{real} axis Im$\,\Lambda =0$.
In this case, all coefficients of the characteristic polynomial $P(w, \Lambda)$ are real, and therefore either the three zeros
$w_1(\Lambda)$, $w_2(\Lambda)$, $w_3(\Lambda)$ are real, or one is real and two are complex conjugate.
In the first case the three wave numbers $k_j$ are real, while in the second case none of them is real, which leads to:
\begin{proposition}
\label{prop:1}
If $\Lambda$ is real then it  belongs to the spectrum $ \mathbb{S}^{\Lambda}$ if and only if the $w$-discriminant of the polynomial (\ref{PwLambda}) is non-negative, namely if $\Delta_w\, P(w, \Lambda)\geq 0$, where
\begin{equation}
\label{DISw}
\Delta_w\, P(w, \Lambda)=k_1^2 \,k_2^2 \,k_3^2 = 4 \Lambda^3 -8q^2 \Lambda^2 +4q (q^3-9r)\Lambda -27 r^2+4rq^3 \;\;.
\end{equation}
\end{proposition}
Thus, as shown by this expression (\ref{DISw}), the large and positive real values of $\Lambda$ do belong to $\mathbb{S}^{\Lambda}$, while the large and negative real values of $\Lambda$ do not.
More explicitly, we approximately solve the equation $P(w, \Lambda)=0$ around the point at infinity of the complex $\Lambda$-plane and obtain
the following asymptotic expressions of the three roots $w_j$
\begin{equation}
\label{largelambda}
\begin{array}{l}
w_1(\Lambda)=\sqrt{\Lambda} -\frac{r}{2\Lambda}  + O(1/\Lambda^{3/2})\\
w_2(\Lambda)=-\sqrt{\Lambda} -\frac{r}{2\Lambda} + O(1/\Lambda^{3/2})\\
w_3(\Lambda)=q +\frac{r}{\Lambda} + O(1/\Lambda^2)\,,
\end{array}
\end{equation}
where the labelling index $j$ is arbitrary.
Consequently, if $\Lambda$ is real, large and positive then also the three wave numbers $k_j(\Lambda)$, see (\ref{k-omega}), are real and large. If instead $\Lambda$ is real and negative, and its modulus is large enough, no real wave number $k_j$ exists and $\Lambda$ does not belong to the spectrum.
If $\Lambda$ is real, large and positive, the $w$-discriminant (\ref{DISw})
is positive, and it remains so while moving $\Lambda$ along the real axis towards the origin, until it reaches its first zero, say $\Lambda_+$, of the discriminant  $\Delta_w P(w, \Lambda)$, see (\ref{DISw}). At this zero, this discriminant generically changes its sign and one of the three wave numbers ${k_j(\Lambda)}$ vanishes.
In turn, the $w$-discriminant (\ref{DISw}) is a cubic polynomial of the variable $\Lambda$ with real coefficients. Therefore, it has either one or three real $\Lambda$-zeros. In the former case, the discriminant $\Delta_w P(w, \Lambda)$ is negative for all values $\Lambda < \Lambda_+$ and hence, according to our Proposition \ref{prop:1}, the real part of the spectrum $\mathbb{S}^{\Lambda}$ is the semi-axis $\Lambda_+ \leq \Lambda < +\infty$. If instead the discriminant $\Delta_w P(w, \Lambda)$ has three real zeros, say $\Lambda_0< \Lambda_- < \Lambda_+$, then the interval $\Lambda_- < \Lambda < \Lambda_+$ cannot belong to the spectrum since in this interval the discriminant $\Delta_w P(w, \Lambda)$ is negative (see Proposition \ref{prop:1}).
Thus, this interval is a finite gap of the spectrum. Indeed, in this case the real part of the spectrum consists on the finite interval $\Lambda_0< \Lambda < \Lambda_-$ where the discriminant $\Delta_w P(w, \Lambda)$ is positive, and the semi-axis $\Lambda_+ \leq \Lambda < +\infty$.
This finite gap should be considered as a distinctive feature of the spectra in our classification.
Indeed, the $\Lambda$-discriminant of the discriminant $\Delta_w P(w, \Lambda)$, that is $\Delta_{\Lambda} \Delta_w P(w, \Lambda)$, depends only on the two parameters $q$, $r$ and serves our classification purpose, as summarised by the following
\begin{proposition}
\label{prop:2}
Let $\Delta_\Lambda\Delta_wP(w, \Lambda)$ be the $\Lambda$-discriminant of the discriminant (\ref{DISw}), that is
\begin{equation}
\label{DISDIS}
\Delta_\Lambda \Delta_w P(w, \Lambda)= 16r(8q^3 -27r)^3\,.
\end{equation}
The spectrum $\mathbb{S}^{\Lambda}$ has one, and only one, finite gap (G) on the real axis if and only if $\Delta_\Lambda \Delta_w P(w, \Lambda)>0$, namely, if and only if $r(8q^3 -27r)>0$, and it has no gap if $r(8q^3 -27r)<0$. The gap opening and closing threshold values of the parameters are $r=0$ and $r=(8/27)\,q^3$.
\end{proposition}
Let us now turn our attention to the complex values of $\Lambda$ that are not on the real axis and yet belong to the spectrum, $\Lambda \in \mathbb{S}^{\Lambda}$, Im$(\Lambda)\neq 0$.
To this purpose, it is far more convenient to introduce the new polynomial
\begin{subequations}
\label{polyCharW}
\begin{equation}
\label{Pz}
\mathcal{P}(\zeta,\Lambda)= (\zeta-k^2_1)(\zeta-k^2_2)(\zeta-k^2_3)=\zeta^3 +\gamma_2 \, \zeta^2 +\gamma_1 \,\zeta + \gamma_0\,,
\end{equation}
whose roots $\zeta_j(\Lambda)$, $\,j=1, 2, 3$, are the squares of the differences of the $\Lambda$-dependent roots $w_j(\Lambda)$ of the polynomial $P(w, \Lambda)$, see (\ref{k-omega}),
\begin{equation}
\label{zroots}
\zeta_j(\Lambda)= k^2_j(\Lambda)=(w_{j+1}-w_{j+2})^2\,,\quad j=1,\,2,\,3\,\;\textrm{mod}\,3\,.
\end{equation}

The coefficients of the polynomial $\mathcal{P}(\zeta,\Lambda)$ (\ref{Pz}) can be computed explicitly in terms of the coefficients of the polynomial $P(w, \Lambda)$ (\ref{PwLambda}) (see \cite{degasperis2019}), and read
\begin{equation}
\label{coeffPz}
\begin{array}{l}
\gamma_2 =- 2(3 \Lambda+q^2)\,,\\
\gamma_1 = {(3 \Lambda + q^2)}^2\,,\\
\gamma_0= - 4 \Lambda^3 +8q^2 \Lambda^2 -4q (q^3-9r)\Lambda +27 r^2-4rq^3\,.
\end{array}
\end{equation}
\end{subequations}
The definition (\ref{Pz}) shows that the $w$-discriminant (\ref{DISw}) of the polynomial $P(w, \Lambda)$ is simply related to this new polynomial (\ref{Pz}) as it reads
\begin{equation}
\label{discrPw}
\Delta_wP(w, \Lambda) = -\mathcal{P}(0,\Lambda)\,.
\end{equation}
This property allows one to rewrite our Proposition (\ref{prop:2}) on gap characterisation in terms of $\mathcal{P}(0,\Lambda)$.
The spectrum $\mathbb{S}^{\Lambda}$ itself can be redefined as the locus of the $\Lambda$-zeros of $\mathcal{P}(\zeta,\Lambda)$ which correspond to non-negative values of the variable $\zeta$, say $\zeta\geq 0$.
Thus, to each $\zeta\geq 0$ there correspond three values of $\Lambda$ which belong to $\mathbb{S}^{\Lambda}$. In particular, since this polynomial $\mathcal{P}(\zeta,\Lambda)$ also takes the expression
\begin{equation}
\label{PzL}
\begin{array}{l}\mathcal{P}(\zeta,\Lambda)=-4\left[\Lambda-\Lambda_1(\zeta)\right] \left[\Lambda-\Lambda_2(\zeta)\right] \left[\Lambda-\Lambda_3(\zeta)\right]=\\
\\
=-4\Lambda^3 + \Lambda^2(9\zeta +8q^2)-2\Lambda(3\zeta^2-3q^2\zeta+2q^4-18qr)+\zeta^3-2q^2\zeta^2+q^4\zeta+27r^2-4q^3 r,
\end{array}
\end{equation}
with real coefficients for real $\zeta$, we conclude that
\begin{proposition}
\label{prop:3}
The spectrum $\mathbb{S}^{\Lambda}$ is symmetric with respect to the the real axis.
\end{proposition}
Indeed, if $\Lambda$ and $\Lambda^\ast$ are roots of $\mathcal{P}(\zeta,\Lambda)$ for a given non-negative $\zeta$, then both $\Lambda$ and  $\Lambda^\ast$ belong to the spectrum.
Thus the spectrum $\mathbb{S}^{\Lambda}$ is a symmetric piecewise smooth curve in the complex $\Lambda$-plane.
 The transition from a triplet of real $\Lambda$-roots of $\mathcal{P}(\zeta,\Lambda)$ to a pair of complex conjugate $\Lambda$-roots and one real $\Lambda$-root (or viceversa) originates from a collision of two real (or two complex conjugate) $\Lambda$-roots, while changing the value of the variable $\zeta$.
This happens at a zero $\zeta_j$ of the discriminant
\begin{subequations}
\begin{equation}\
\label{Qz}
Q(\zeta)
=\Delta_\Lambda \mathcal{P}(\zeta,\Lambda)\,,
\end{equation}
namely, if $Q(\zeta_j) =0$, provided this discriminant changes its sign.
This discriminant $Q(\zeta)$ turns out to be of fifth degree. However, it factorises as
\begin{equation}
\label{Q1Q2}
\begin{array}{c}
Q(\zeta) =4\,Q_1^2(\zeta)\,Q_2(\zeta)\\
\\
Q_1(\zeta)= 18q\zeta + 27r -8q^3,\quad  Q_2(\zeta)=\zeta^3 -8q^2\zeta^2 + 8q(2q^3 -9r)\zeta + 4r(8q^3 -27r)\,.
\end{array}
\end{equation}
\end{subequations}
After the collision where the discriminant $Q(\zeta)$ changes its sign, the two $\Lambda$-roots scatter off the real axis.
The change of sign makes the factor $Q_1^2(\zeta)$, see (\ref{Q1Q2}), irrelevant to this analysis, so that hereafter we focus our attention on the factor $Q_2(\zeta)$ only.
This polynomial has generically three simple roots,
\begin{equation}
\label{factQ2}
Q_2(\zeta)=(\zeta-\zeta_1)(\zeta-\zeta_2)(\zeta-\zeta_3)\,,
\end{equation}
which depend only on the parameters $q$, $r$, and may be either all three real, if the discriminant of $Q_2(\zeta)$ is positive,
\begin{subequations}
\label{DQ2}
\begin{equation}
\label{DQ2P}
\Delta_{\zeta} Q_2(\zeta)=16r{\left(16q^3-27r\right)}^3 >0\,,\quad\zeta_1 < \zeta_2 < \zeta_3\,,
\end{equation}
or two of them are complex conjugate and one is real, if this discriminant is instead negative,
\begin{equation}
\label{DQ2M}
\Delta_{\zeta} Q_2(\zeta)=16r{\left(16q^3-27r\right)}^3 <0\,,\quad\zeta_1=\mu+i\rho\,,\quad \zeta_2 =\mu-i\rho\,,\quad \text{Im}(\zeta_3)=0\,,
\end{equation}
\end{subequations}
where $\mu$, $\rho$ and $\zeta_3$ are real.
Thus, our classification is mainly based on the function $Q_2(\zeta)$ and on its discriminant $\Delta_{\zeta} Q_2(\zeta)$.
In this context, the necessary existence condition of a gap, which follows from our Proposition (\ref{prop:2}), can also be established as $Q_2(0)=
-\zeta_1\zeta_2\zeta_3= 4r(8q^3 -27r) >0$.
In the following classification, we find it more convenient to play with the sign of $qr$ rather than with the sign of $q$ and of $r$ separately. Thus, for instance, the existence condition of a gap of the spectrum reads $(qr)\left[8q^4-27(qr)\right]>0$ which is never satisfied if $qr<0$ and it is satisfied only for $0<qr<\left(8/27\right)q^4$.

Consider first a spectrum $\mathbb{S}^{\Lambda}$ whose parameters $q$, $r$ satisfy the positive discriminant inequality (\ref{DQ2P}), namely $(qr)\left[16q^4-27(qr)\right]>0$.  In this case, for large enough values of positive $\zeta$, namely for $\zeta_3<\zeta<+\infty$, as well as for $\zeta_1<\zeta<\zeta_2$, the function $Q_2(\zeta)$ is positive,
$Q_2(\zeta)>0$, while $Q_2(\zeta)<0$ for $\zeta_2<\zeta<\zeta_3$ and for $-\infty<\zeta<\zeta_1$. In order to look for distinctive features of the spectrum, we have to analyse the following four different cases separately according to the positiveness of the three $\zeta$-roots of $Q_2(\zeta)$.
\begin{enumerate}
\item
$\zeta_1 < \zeta_2 < \zeta_3<0$: this case (all roots are negative) is excluded by the Vieta relation $\zeta_1 + \zeta_2 +\zeta_3=8q^2$.
\item
$\zeta_1 < \zeta_2 <0< \zeta_3$: the Vieta relation $\zeta_1 \zeta_2 \zeta_3 =4r(27r-8q^3)$, implies $(qr)\left[27(qr)-8q^4\right]>0$,  while the positive  discriminant condition (\ref{DQ2P}), $(qr)\left[16q^4-27(qr)\right] >0$, does not allow $qr$ to be negative.
We conclude that only the interval $(8/27)q^4<qr<(16/27)q^4$ is allowed. In this case, two $\Lambda$-roots collide for $\zeta=\zeta_3$ and, for $0\leq \zeta<\zeta_3$, the spectrum acquires two complex conjugate curves in the $\Lambda$-plane.
We refer to this complex part of the spectrum as branch (B), since the two end-points at $\zeta=0$ of the two $\Lambda$-roots trajectories  do not generically coincide with each other (see Figure \ref{fig:Ba} below). No gap is possible and we term this spectrum type 0G 1B 0L, or of B-type.
\item
$\zeta_1 <0< \zeta_2 < \zeta_3$: the same Vieta relation used above necessarily requires $0<qr<(8/27)q^4$, which is also compatible with the positive discriminant condition.  Two $\Lambda$-roots collide for $\zeta=\zeta_3$, get off the real axis and collide again for $\zeta=\zeta _2$ on the real $\Lambda$-axis thereby forming one complex closed curve, which we term loop (L) (see Figure \ref{fig:LGa} below). In this case there is no branch.
However, a gap exists because its existence condition $(qr)\left[8q^4-27(qr)\right]>0$, see our Proposition (\ref{prop:2}), is satisfied.
Thus, we call this spectrum of type 1G 0B 1L, or of LG-type.
\item
$0<\zeta_1 < \zeta_2 < \zeta_3$: this case is not allowed by two Vieta relations, which lead to the inequalities $6q^4-27(qr)>0$ and $(qr)\left[27(qr)-8q^4\right]>0$, and by the positive discriminant condition (\ref{DQ2P}).
\end{enumerate}
Let us consider now those spectra $\mathbb{S}^{\Lambda}$ whose parameters $q$, $r$ satisfy the negative discriminant inequality (\ref{DQ2M}) instead, namely $(qr)\left[16q^4-27(qr)\right]<0$.
Again we distinguish all possible cases according to the $\zeta$-roots (\ref{DQ2M}).
\begin{enumerate}
\item
$\zeta_3<0$: no spectrum exists because the Vieta relations for the polynomial $Q_2(\zeta)$, see (\ref{Q1Q2}), and the sign of the discriminant (\ref{DQ2M}) are never compatible with each other.
\item
$\zeta_3>0$, $\mu<0$: the Vieta relations and the discriminant sign imply that the spectrum exists if either $qr<0$ or $(16/27)q^4<qr$. This spectrum is of type 0G  1B  0L.
\item $\zeta_3>0$, $\mu>0$: again, combining Vieta relations and the discriminant negative sign shows that the spectrum exists only if $qr<0$, and it is of type 0G 1B  0L.
\end{enumerate}

These observations above conclude our characterisation and classification of spectra, which are summarised in the following Table \ref{table_summary}.
\begin{table}[h!]
\caption{Stability spectra\label{table_summary}}
\begin{center}
\begin{tabular}{ccc}
$qr<0\qquad$ & $0<qr<\frac{8}{27}q^4\qquad$ & $\frac{8}{27}q^4 <qr$\\
\hline
0G 1B 0L$\qquad$ & 1G 0B 1L$\qquad$ & 0G 1B 0L
\end{tabular}
\end{center}
\vspace*{-4pt}
\end{table}
As stated after our Definition \ref{def:Spectrum} of the spectrum $\mathbb{S}^{\Lambda}$, see also (\ref{q1}), the classification as in Table \ref{table_summary} has been derived with the assumption that $q\neq0$.
For the sake of completeness, we now remove this limitation by computing the spectrum for $q=0$ with $r\neq 0$ (see below for $r=0$).
We first observe that in this special case  the polynomial $\mathcal{P}(\zeta,\Lambda)$ reads
\begin{equation}
\label{q0poly}
\mathcal{P}(\zeta,\Lambda)=-\left(4\Lambda-\zeta\right)\left(\Lambda-\zeta\right)^2 +27 r^2\,,
\end{equation}
and therefore $\mathcal{P}(0,\Lambda)=-4\Lambda^3 +27 r^2$.
According to our Proposition (\ref{prop:1}) and to the relation (\ref{discrPw}), the real part of the spectrum is the semi-axis
$\Lambda_0\leq\Lambda<+\infty$ with $\Lambda_0=\left[(27/4)r^2\right]^{1/3}$.
As for the complex part of the spectrum, we note that $Q_2(\zeta)=\zeta^3-108 r^2$, see (\ref{Q1Q2}), and therefore the only real zero of $Q_2(\zeta)$ is the positive number $\zeta_3=(108 r^2)^{1/3}$. For this value of $\zeta$, the polynomial $\mathcal{P}(\zeta,\Lambda)$ must have a double real $\Lambda$-root $\Lambda_B$, which is found to be $\Lambda_B=(1/2)\zeta_3=(1/2)\left(108r^2\right)^{1/3}=2^{1/3} \Lambda_0$.
 As $\zeta$ is moved from $\infty$ towards the origin, two real $\Lambda$-roots collide with each other at $\Lambda_B$ for $\zeta=\zeta_3$ to give rise to a branch, whose off real parts terminate for $\zeta=0$ at the complex conjugated points $\Lambda_0 \exp(2i\pi/3)$, $\Lambda_0 \exp(-2i\pi/3)$. Thus, for $q=0$, $r\neq 0$ the spectrum is of type 0G 1B 0L.

A similar analysis can be carried out at any zero of discriminants of cubic polynomials, in particular to discuss threshold phenomena.
Indeed, the inequalities shown in Table \ref{table_summary} divide the plane $(q,r)$ into regions with different types of spectra.
At the boundaries of these regions, the inequalities in Table \ref{table_summary} turn into equalities which characterise thresholds.
One simple and interesting instance of such thresholds occurs at the special value $r=0$.
In this special case, the spectrum $\mathbb{S}^{\Lambda}$ is readily found because the roots of the characteristic polynomial (\ref{PwLambda}) $P(w,\Lambda)= (w-q)(w^2-\Lambda)$ are explicit.
The corresponding spectrum is entirely real and we will call it of type 0G 0B 0L. Indeed, the wave number functions $k_j(\Lambda)$ have the simple expression
 \begin{equation}
 k_1(\Lambda)=-q-\sqrt{\Lambda}\,,\quad k_2(\Lambda)=q-\sqrt{\Lambda}\,,\quad k_3(\Lambda)=2\sqrt{\Lambda}\,,\quad 0\leq \Lambda < +\infty\,.
 \end{equation}
 A second threshold occurs for $27r-8q^3=0$, where a gap disappears. At these threshold points of the $(q,r)$-plane the discriminant $Q(0)$, see (\ref{Q1Q2}), vanishes with the implication that the polynomial $\mathcal{P}(0,\Lambda)= -4\left[\Lambda +(1/3)q^2\right]^2 \left[\Lambda -(8/3)q^2\right]$ has a double $\Lambda$-root at $\Lambda=-(1/3)q^2$, where a branch closes up and becomes a loop.
The corresponding spectrum can be classified as 1G 0B 1L where however the gap is reduced to just one point.

Let us now turn our attention to the time variable $t$ and to the stability of the plane wave solution (\ref{CW}).
This point requires investigating whether the frequencies $\omega_1(\Lambda)$, $\omega_2(\Lambda)$, $\omega_3(\Lambda)$, see (\ref{k-omega}), are real numbers for $\Lambda \in \mathbb{S}^{\Lambda}$.
Our starting point is their relation to the wave numbers $k_j(\Lambda)$
\begin{equation}
\label{omega}
\omega_j = \frac{1}{3} k_j \left(2q+k_j +2k_{j+1}\right),\quad j=1, 2, 3\;\text{mod}\; 3\,,
\end{equation}
which can be derived from their definition (\ref{k-omega}) by first inverting the map $\{w_j\}\rightarrow \{k_j\}$, that is (see (\ref{k-omega}))
\begin{equation}
\label{wk}
w_j = \frac{1}{3} \left(q+k_j +2k_{j+2}\right),\quad j=1, 2, 3\;\text{mod}\; 3\,.
\end{equation}
This relation (\ref{omega}) is clearly not a dispersion relation since each frequency $\omega_j$ is written in terms of two wave numbers, $k_j$ and $k_{j+1}$.
In order to compute the dispersion relation, we need to eliminate the variable $\Lambda$ among the algebraic relations $P(w, \Lambda)=0$, $\mathcal{P}(\zeta,\Lambda)=0$ and $\omega_j= k_j(q-w_j)$. This requires a tedious but straightforward computation aimed to factorise high degree polynomials. This is done with the help of the additional algebraic relation $\mathcal{R}(\xi,\Lambda)=0$ with the following specifications
\begin{subequations}
\label{polyCharW2}
\begin{equation}
\label{R}
\mathcal{R}(\xi,\Lambda)= (\xi-\omega^2_1)(\xi-\omega^2_2)(\xi-\omega^2_3)=\xi^3 +\delta_2 \xi^2 +\delta_1 \xi + \delta_0  \;\,,
\end{equation}
where the coefficients $\delta_n(\Lambda)$ can be explicitly computed as in \cite{degasperis2018}, and have the following expression
\begin{equation}
\label{coeffRxi}
\begin{array}{l}
\delta_2 =    - 2\left(q^4 - 6qr - 2q^2 \Lambda +\Lambda^2\right),\\
\delta_1 =   \left(q^4 - 6qr - 2q^2 \Lambda + \Lambda^2 \right)^2,\\
\delta_0=   r^2 \left(-4q^3 r + 27r^2 - 4q^4\Lambda +36qr\Lambda + 8q^2\Lambda^2 - 4\Lambda^3\right) \,.
\end{array}
\end{equation}
\end{subequations}
This additional polynomial $\mathcal{R}(\xi,\Lambda)$, as shown in (\ref{R}), is defined in the same way as $\mathcal{P}(\zeta,\Lambda)$ by substituting the wave numbers $k_j$ with the frequencies $\omega_j$, and by replacing the polynomial $P(w, \Lambda)$ with the characteristic polynomial of the matrix $W^2$, that is
\begin{equation}
\label{polyW2}
\det\left[w^2\, \mathbf{1}-W^2 \right]=- P(w, \Lambda) P(-w, \Lambda)=\left(w^2-q^2\right)\left(w^2-\Lambda \right)^2 + 2qr \left(w^2-\Lambda \right) - r^2 \,.
\end{equation}
The upshot of this calculation is the three-branch dispersion relation $H(\pm k_j,\pm\omega_j)=0$, where $H(k,\omega)$ is the polynomial
\begin{equation}
\label{H}
H(k,\omega)= \omega^3-4q k \omega^2+k^2\left(4q^2-k^2\right)\omega -4rk^3\,.
\end{equation}
It is worth noticing that this dispersion relation $H(k,\omega)=0$ can be obtained by means of a standard Fourier approach to the linearised equations. In other words, the continuous wave solution (\ref{CW}) is so special that its stability properties can be investigated by both Fourier expansion and our Lax pair method. However, since the Fourier expansion approach works \textit{only} for this particular solution, we have turned our focus on the integrability methods in order to explicitly apply the mathematical formalism (see Section \ref{sec:integrable}) in a simple context. The next, and harder, task will be to extend the present analysis to other known solutions, such as solitary waves, say solitons, periodic and rogue waves.

Let us now consider the linear stability of the plane wave solution (\ref{CW}) for  \emph{generic} values of $q$ and $r$.
Our analysis has shown that the $\Lambda$-spectra have a real component  and a complex, off real one.
The former lies on the real axis, and at each point all three wave numbers $k_1(\Lambda)$, $k_2(\Lambda)$, $k_3(\Lambda)$ are real.
Since the corresponding frequencies $\omega_j(\Lambda)$, according to their expression (\ref{omega}), are also real, the eigenfunctions (\ref{CWeigenFF}) on this component of the spectrum are bounded in $t$, and do not cause instabilities.
On the contrary, if $\Lambda$  is off the real axis, namely on a branch or on a loop, which are the complex components of $\mathbb{S}^{\Lambda}$, only one of the three wave numbers, for instance $k_3(\Lambda)$, is real, while the other two $k_1(\Lambda)$, $k_2(\Lambda)$ are non-real. On branches and loops, while $ k_3(\Lambda)$ is real, the corresponding frequency $\omega_3(\Lambda)$ is necessarily complex with a non-vanishing imaginary part. This follows from the expression (\ref{omega}). Precisely, since $\omega_3 = \frac{1}{3} k_3\,\left(2q+k_3 +2k_1\right)$, the real frequency is
\begin{equation}
\label{freq3}
\text{Re}(\omega_3)=\frac{1}{3} k_3\left(2q+k_3\right) +\frac23 k_3\, \text{Re}(k_1)\,.
\end{equation}
The corresponding instability is characterised by the gain function (\ref{gain})
\begin{equation}
\label{gain3}
\Gamma_3(\lambda)= \left| \text{Im}(\omega_3) \right|=\frac{2}{3}\,\left| k_3 \,\text{Im}(k_1) \right| \,,
\end{equation}
which is non-vanishing as far as $\Lambda$ remains on the non-real component of the spectrum.
We conclude this analysis with the following
\begin{proposition}
\label{prop:4}
All stability spectra $\mathbb{S}^{\Lambda}$ are classified with respect to the parameters $q$ and $r\neq0$ in two types: the B-type (having a real part and one branch), and the LG-type (having a real part with one finite gap, and one loop). Only for $r=0$ the spectrum is totally real with no complex part.
The continuous wave solution is linearly stable if and only if $r=0$, with $\omega_1=k_1^2+2qk_1$, $\omega_2=-k_2^2+2qk_2$, $\omega_3=0$, and is unstable for all values of $q$ and all non-vanishing values of $r$.
\end{proposition}
Examples of these two types of spectra have been numerically computed, by evaluating the $\Lambda$-roots of $\mathcal{P}(\zeta,\Lambda)$ in (\ref{q0poly}), as $\zeta$ varies from 0 to the largest, positive, real $\zeta$-root of $Q_{2}(\zeta)$ in (\ref{Q1Q2}), as detailed in Appendix C of \cite{degasperis2018}.
In Figures \ref{fig:B}, we show the B-type spectrum (Fig. \ref{fig:Ba}); the corresponding gain function $\Gamma$ on the branch (Fig. \ref{fig:Bb}), which proves that this instability is of baseband type, namely waves are unstable $|k|=0$; the functions $k_j(\Lambda)$, for $j=1,2,3$, if $\Lambda$ is real, together with the function $k_3(\text{Re}(\Lambda))$ on the branch (Fig. \ref{fig:Bc}); the real frequency on the branch as function of $k=k_3$ (Fig. \ref{fig:Bd}).
The same functions are plotted in Figures. \ref{fig:LGb}, \ref{fig:LGc} and \ref{fig:LGd}, for an LG-type spectrum, which is shown in Figure \ref{fig:LGa}. In particular, Figure \ref{fig:LGb} shows that in this case the instability is of passband type, namely, waves are stable for sufficiently small values of $|k|$.

\begin{figure}[h!]
\centering
\begin{subfigure}[b]{0.45\textwidth}
\includegraphics[height=5.5cm]{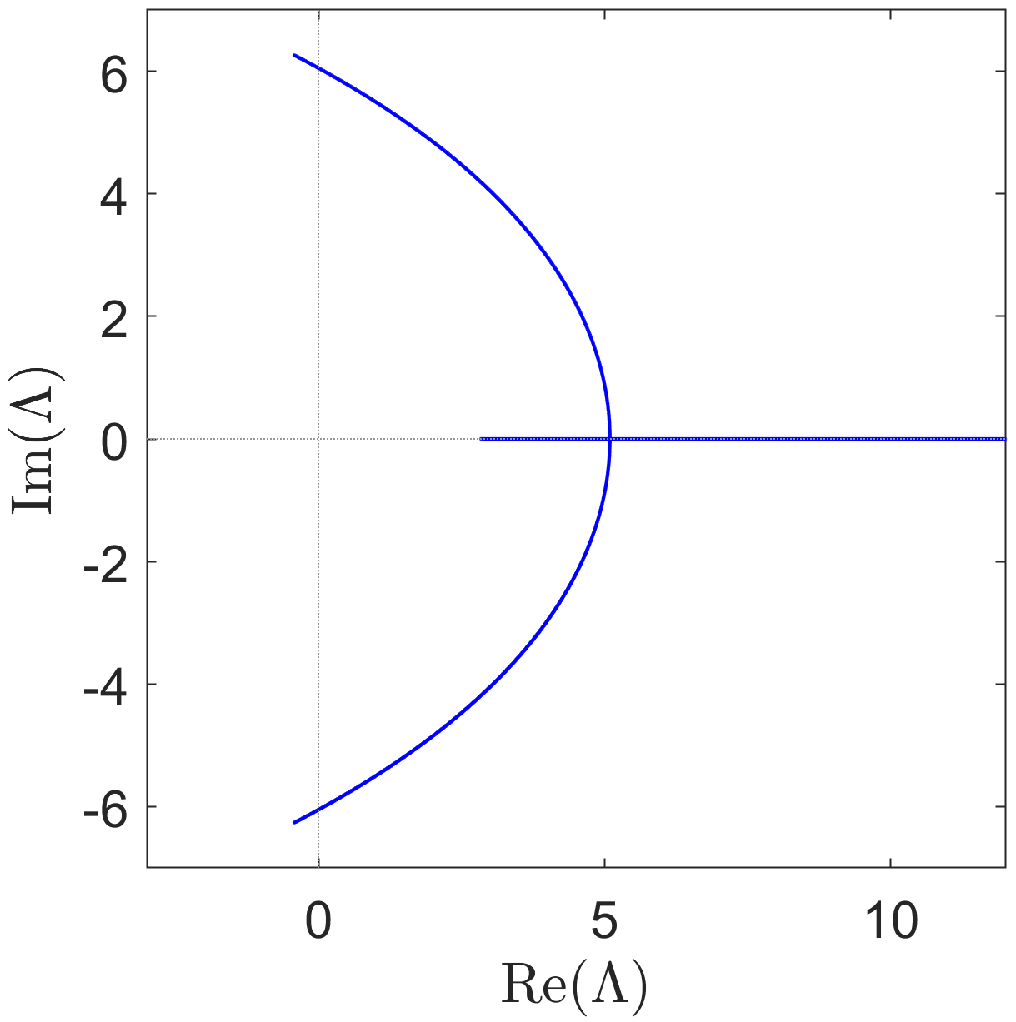}
\caption{Spectrum $\mathbb{S}^{\Lambda}$.\label{fig:Ba}}
\end{subfigure}
\hspace{0.6cm}
\begin{subfigure}[b]{0.45\textwidth}
\includegraphics[height=5.5cm]{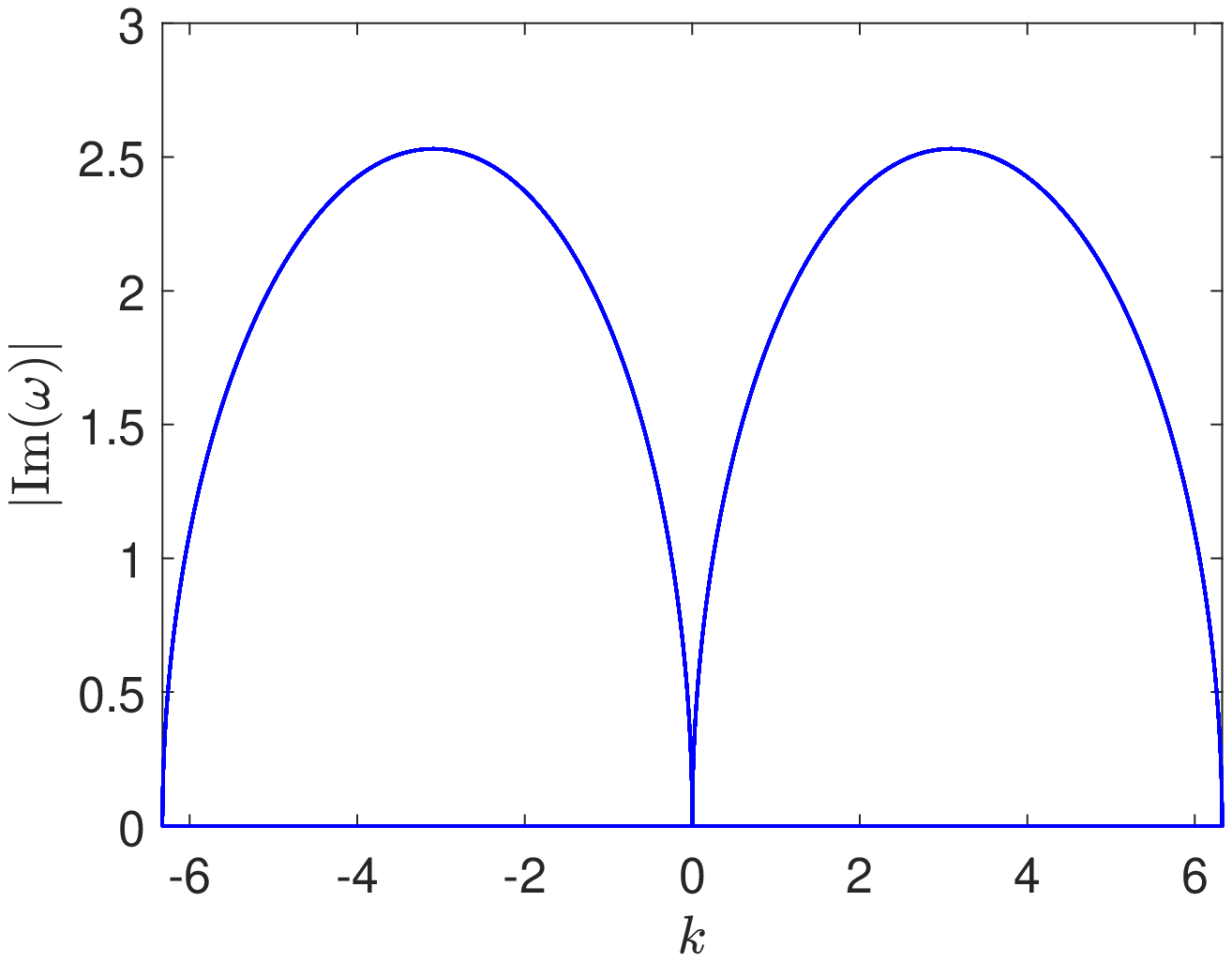}
\caption{Gain function $\Gamma$ versus $k_3$.\label{fig:Bb}}
\end{subfigure}
\begin{subfigure}[b]{0.45\textwidth}
\includegraphics[height=5.5cm]{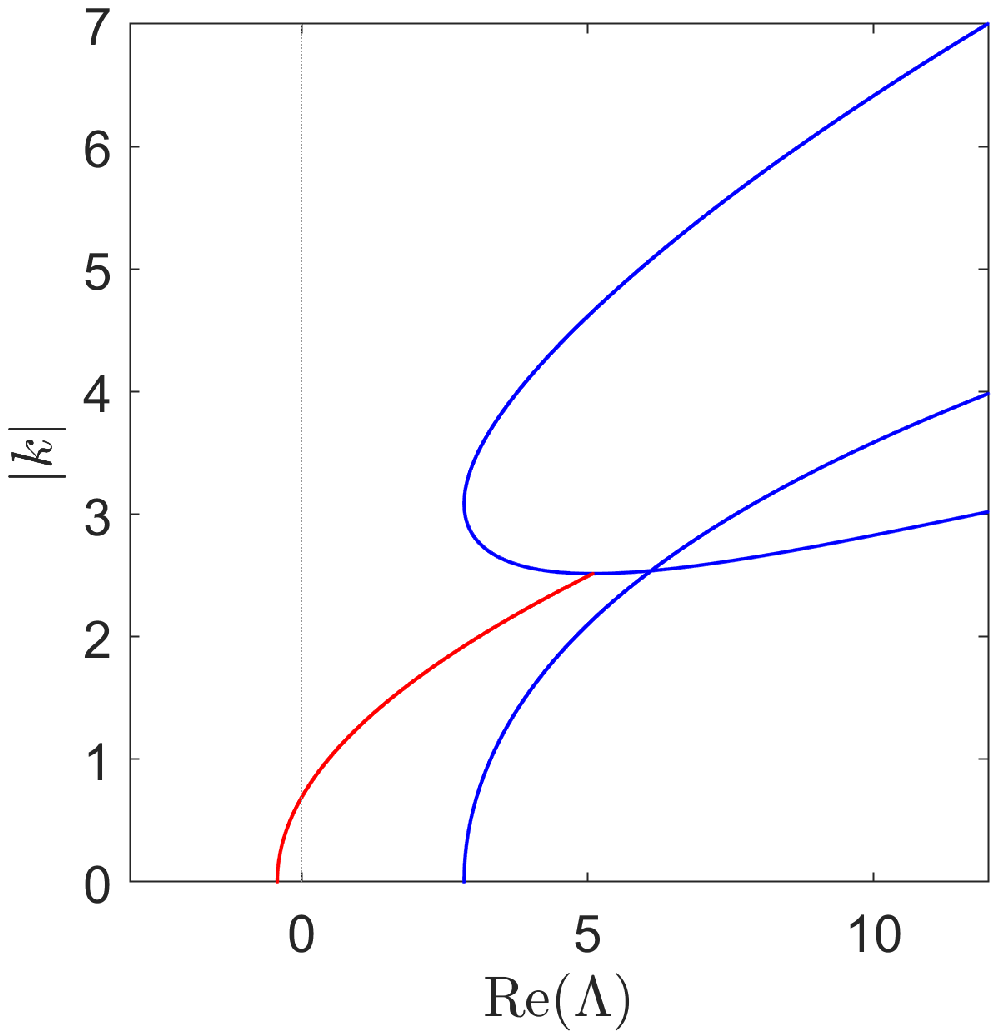}
\caption{Wave numbers versus real (blue) and non-real (red) values of $\Lambda$.\label{fig:Bc}}
\end{subfigure}
\hspace{0.6cm}
\begin{subfigure}[b]{0.45\textwidth}
\includegraphics[height=5.5cm]{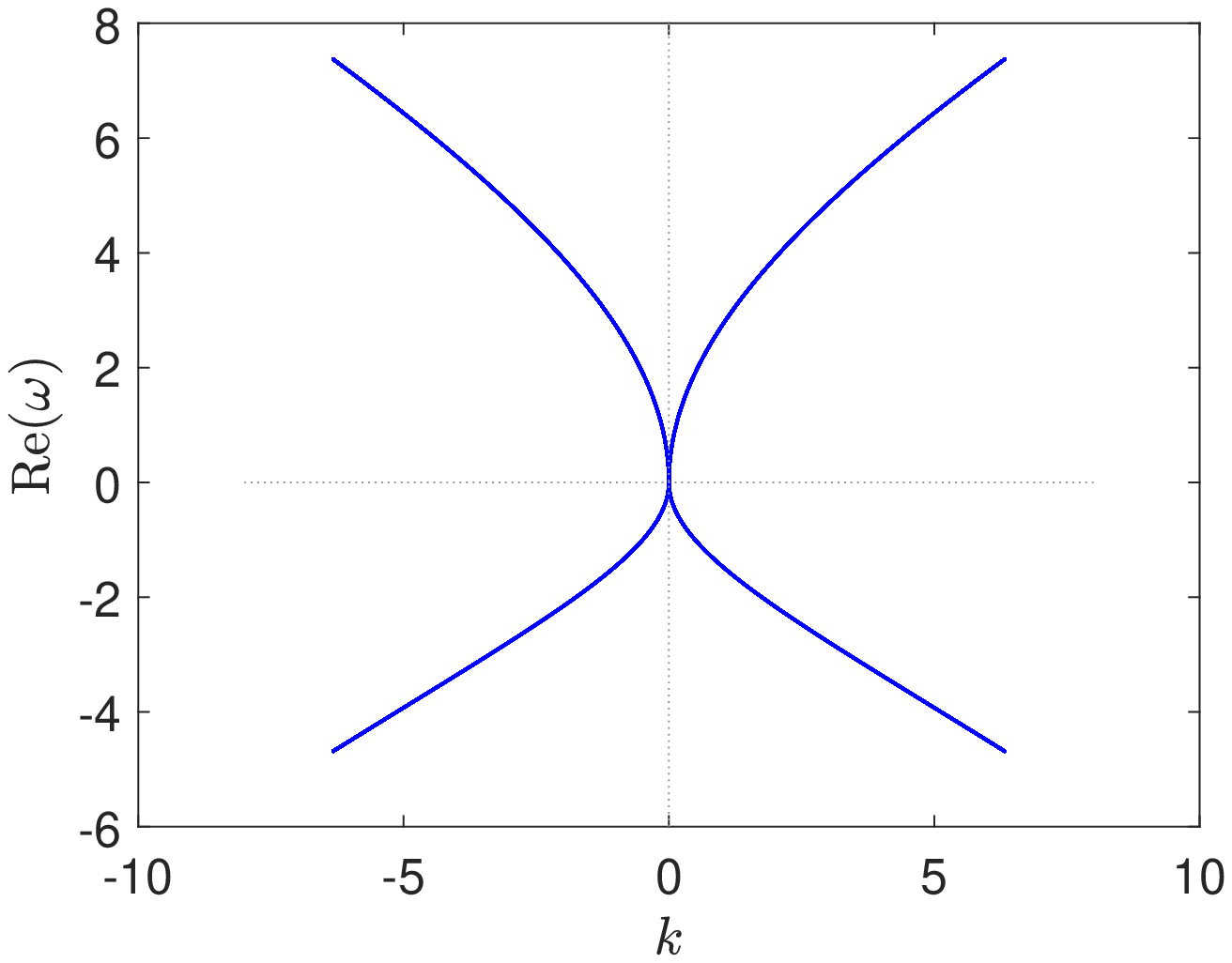}
\caption{Real frequency versus $k_3$ (see (\ref{freq3})).\label{fig:Bd}}
\end{subfigure}
\caption{B-type spectrum, $q=1$, $r=-4$. \label{fig:B}}

\end{figure}
\begin{figure}[h!]
\centering
\begin{subfigure}[b]{0.45\textwidth}
\includegraphics[height=5.5cm]{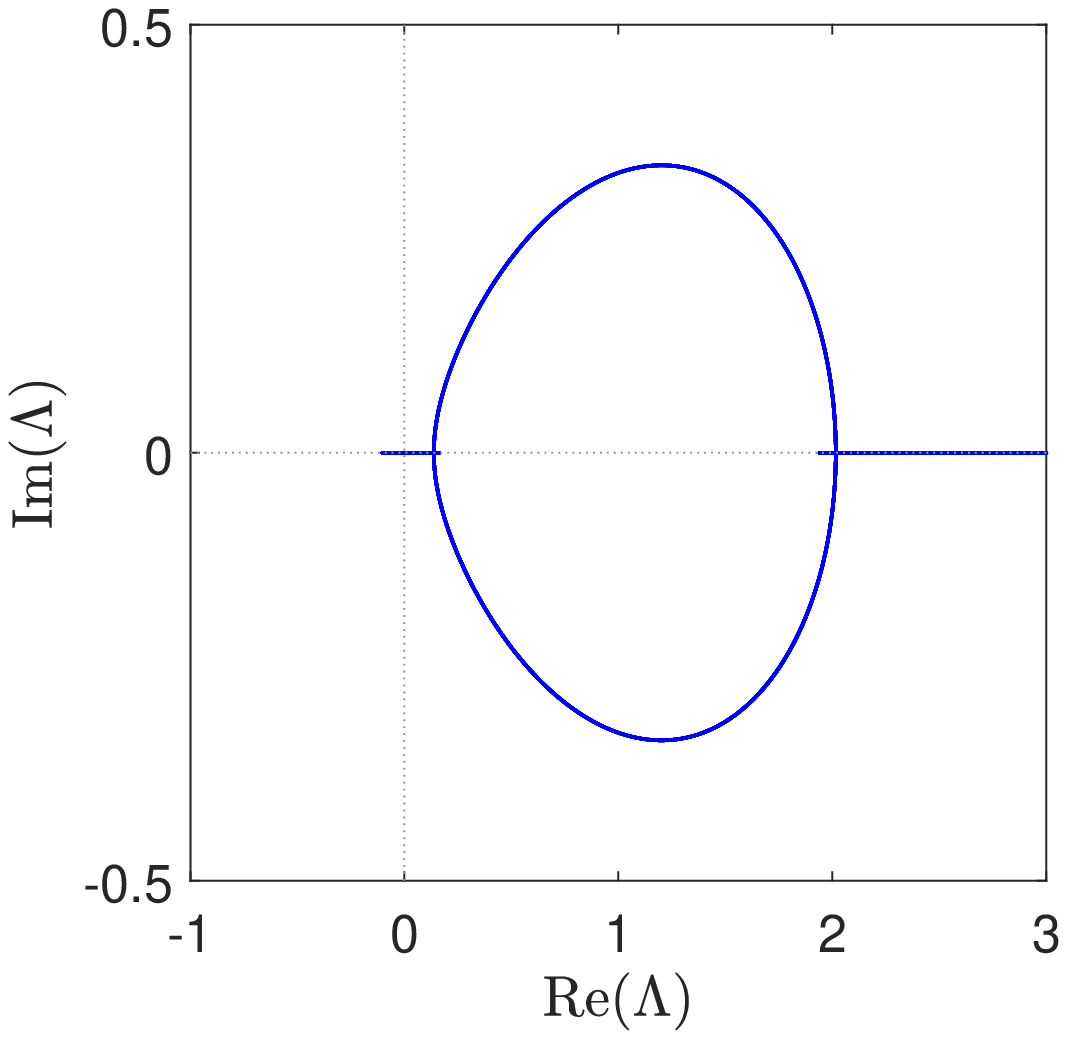}
\caption{Spectrum $\mathbb{S}^{\Lambda}$.\label{fig:LGa}}
\end{subfigure}
\hspace{0.6cm}
\begin{subfigure}[b]{0.45\textwidth}
\includegraphics[height=5.5cm]{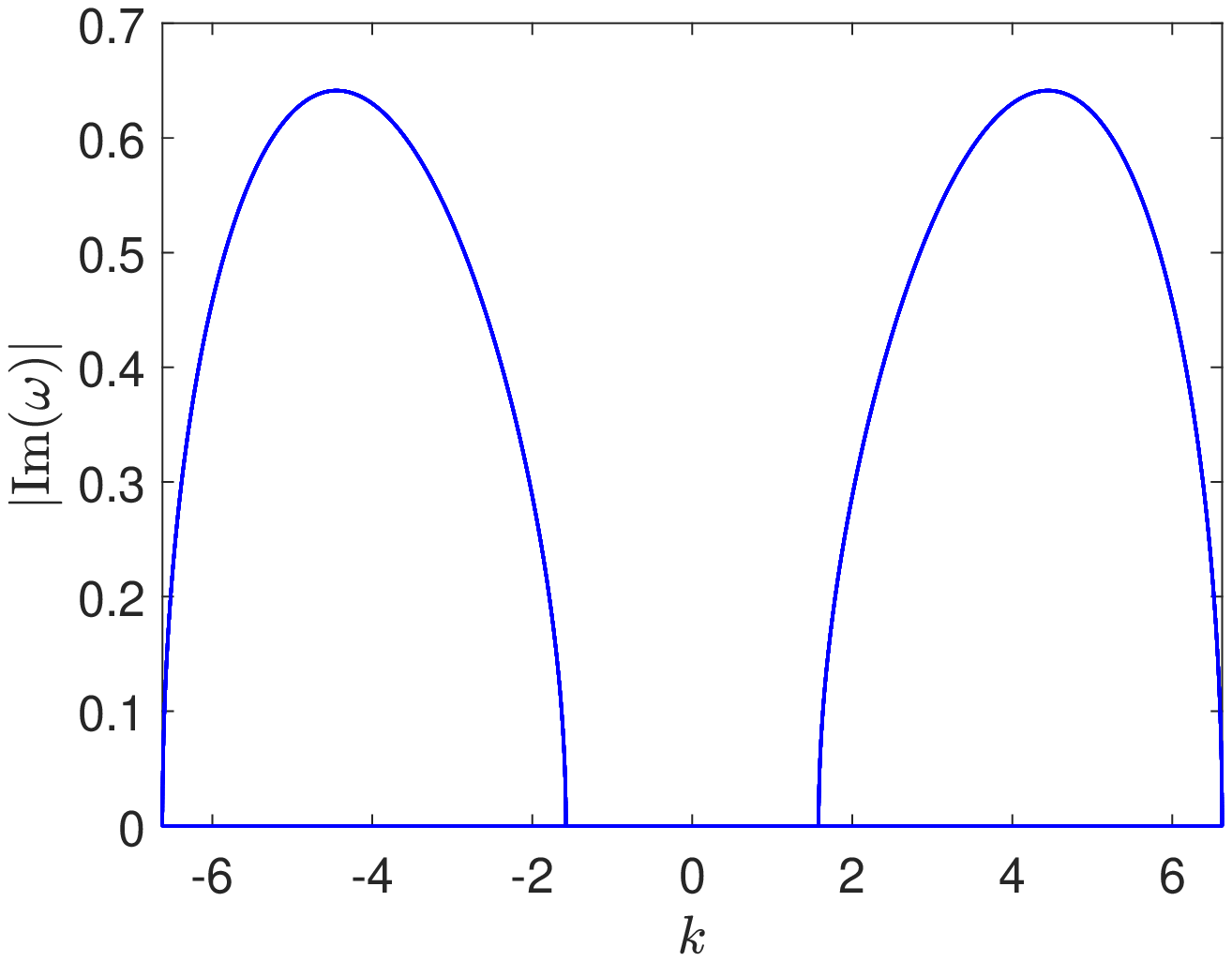}
\caption{Gain function $\Gamma$ versus $k_3$. \label{fig:LGb}}
\end{subfigure}
\begin{subfigure}[b]{0.45\textwidth}
\includegraphics[height=5.5cm]{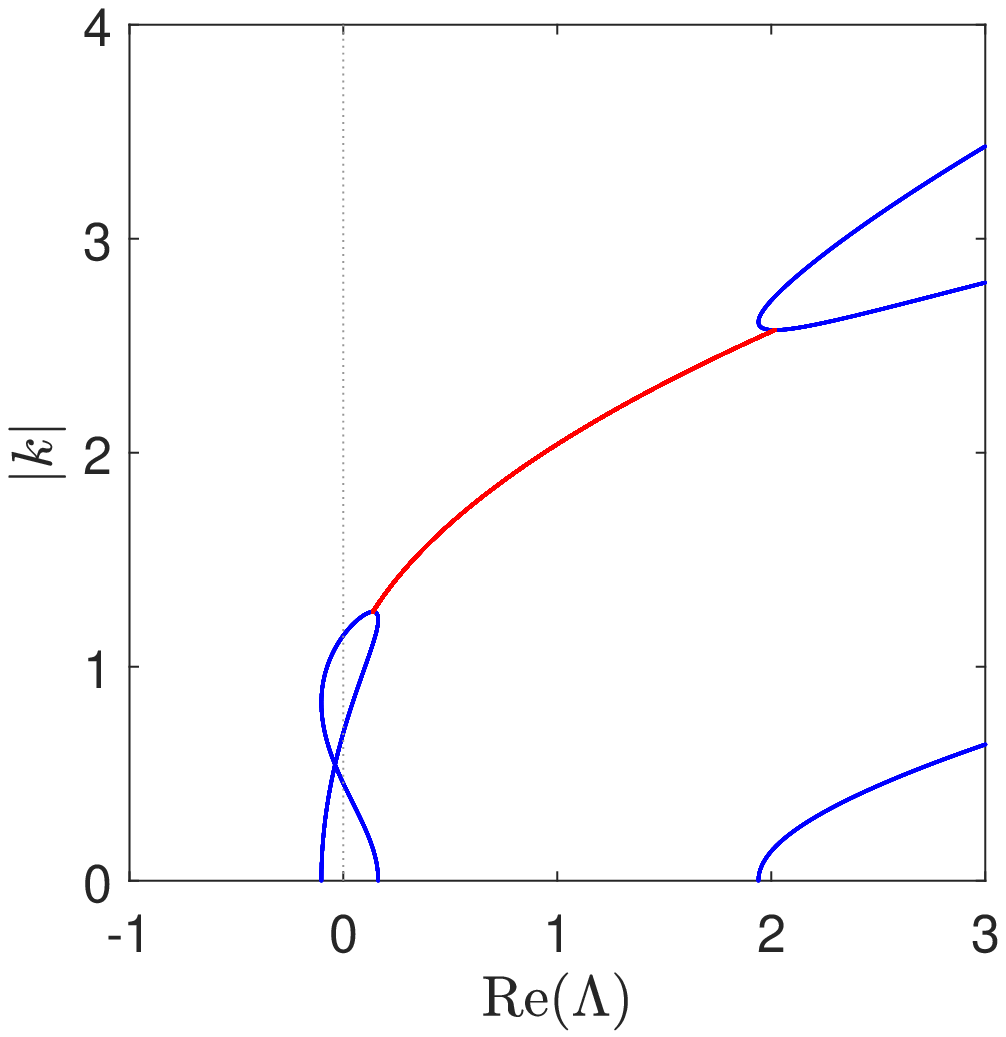}
\caption{Wavenumbers versus real (blue) and non-real (red) values of $\Lambda$.\label{fig:LGc}}
\end{subfigure}
\hspace{0.6cm}
\begin{subfigure}[b]{0.45\textwidth}
\includegraphics[height=5.5cm]{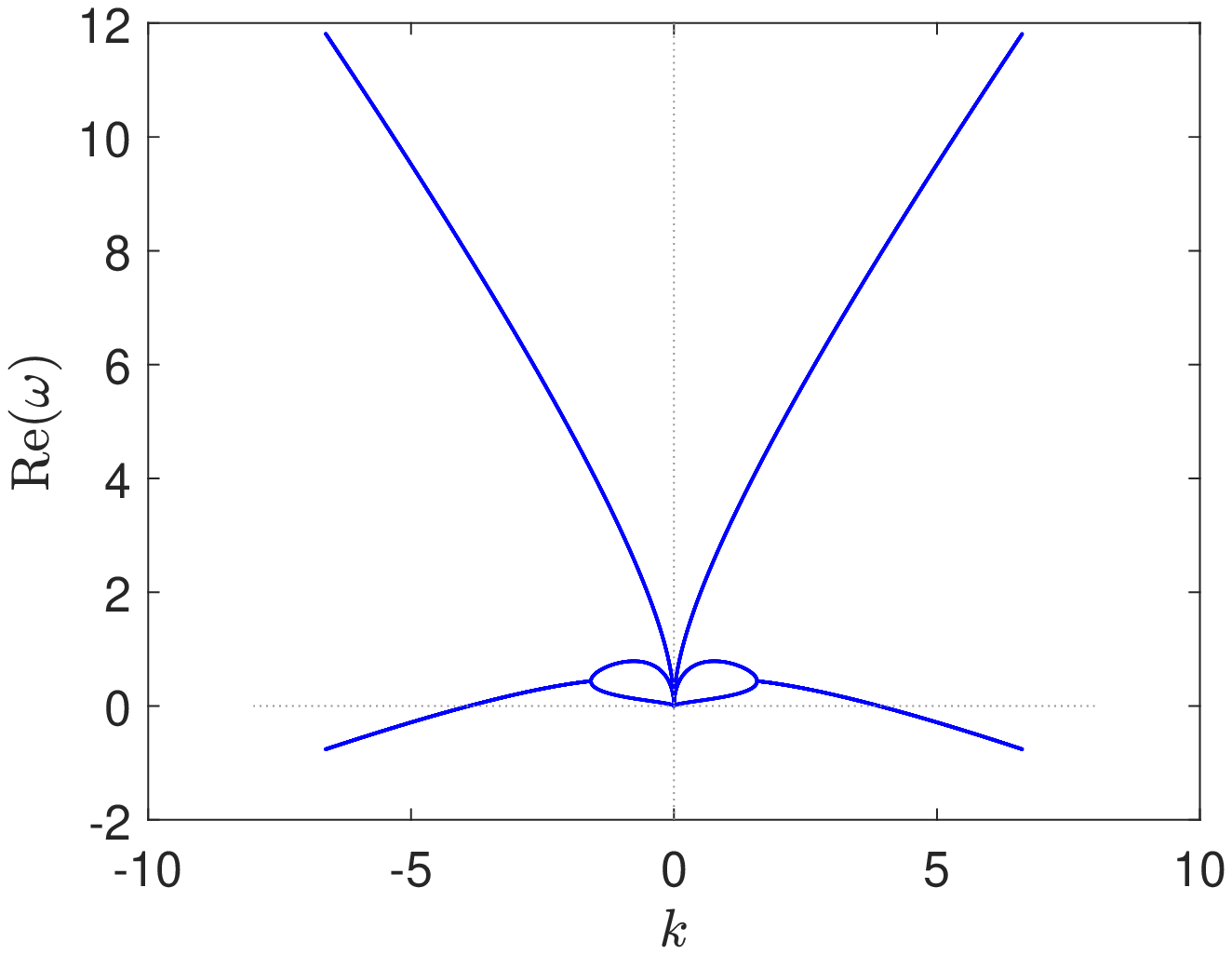}
\caption{Real frequency versus $k_3$ (see (\ref{freq3})).\label{fig:LGd}}
\end{subfigure}
\caption{LG-type spectrum, $q=1$, $r=0.1$. \label{fig:LG}}
\end{figure}

\section{Summary and conclusions}
\label{sec:conclusions}
Research on instabilities of nonlinear waves has witnessed a renewed interest in recent years, following the study of the modulational instability of continuous wave solutions of the focusing NLS equation.
We exploited integrability to construct a method to predict whether a nonlinear wave, described by an integrable, nonlinear system, is linearly stable against small perturbations. Here, we displayed how this approach works in the resonant coupling of long and short waves.
As a side result, we first introduce a new model equation which combines two very well known integrable models, namely the Yajima-Oikawa equation (\ref{YO}) and the Newell equation (\ref{N}), into the more general YON equation (\ref{YON}), which features two arbitrary, real parameters and thus,  it is likely to fit a broader range of physical contexts. This outcome is similar to the one that proves that the Korteweg-de Vries (KdV) and the modified-KdV equations are just two particular cases of the Gardner equation \cite{gardner1968,calogero1982}.

A common feature of all these long wave-short wave equations is that a long wave always arises as generated by short waves. This process, which is guessed by observing the evolution equations (\ref{YON}) themselves, is revealed by the first two conservation laws $\rho_{nt}=g_{nx}$,  $n=1, 2$, where $\rho_{n}$ is the density and $g_{n}$ is the current. The first conservation law is the long wave equation itself, with $\rho_1=L$ and $g_1=2|S|^2 $. The second conservation law turns out to have the conserved density $ \rho_2=|S|^2 -(\alpha/2)L^2$ and the current $g_2=2\,\text{Im}(SS^\ast_x)-2\alpha L |S|^2$.

In this work, we give the $3\times3$ Lax pair associated to the new YON model and display how this Lax pair allows us to construct the basic solutions, or eigenfunctions, of the linearised model equations. These computations are specialised to the continuous wave (or plane-wave) solutions (\ref{CW}) of the YON equations. In this case, the explicit construction of the eigenfunctions of the linearised wave equations allows us to define, and compute, their associated wave numbers $k$ and frequencies $\omega$.
Not only are these physical quantities related to each other by an explicit or implicit dispersion relation (\ref{H}), as in standard Fourier analysis, but they are also parametrically related as $k=k(\Lambda)$, $\omega=\omega(\Lambda)$, where the variable $\Lambda$ lives in the complex plane. In turn, the complex variable $\Lambda$ is simply connected to the spectral variable $\lambda$, which appears in the Lax pair.
The key point is the definition of the \emph{stability spectrum}, an algebraic curve whose geometrical properties convey relevant physical information, and which generically differs from the Lax spectrum. Eigenfunctions, wave numbers and frequencies are defined on the stability spectrum.
In particular, on the real $\Lambda$ part of the spectrum the eigenfunctions are bounded functions of time (stability), while on the non-real part of the spectrum the eigenfunctions exponentially grow with time (instability). This strictly complex part of the spectrum brings its own information: it displays one open branch, if the instability is of baseband type, or a closed loop, if the instability is instead of passband type.

All the spectra are classified by their dependence on essentially one real parameter $r$, see (\ref{pr}). This is a simple function of the amplitudes $a$ and $b$ of the short and, respectively, long waves, of the wave number $q$ of the short wave, and of the self- and cross-coupling constants $\alpha$ and $\beta$ of the model. For a generic choice of all these parameters, only two types of spectra, and therefore of perturbed wave behaviours, emerge from our analysis. Plots illustrate these two typical stability spectra. Their properties show that all values of the parameters lead to instabilities, with the only exception of the case $r=0$, which instead guarantees stability since the associated spectrum has neither a branch nor a loop. This special case corresponds to more than one choice of the physical parameters, according to the expression (\ref{pr}) $r=a^2[2\alpha(q+\alpha b)-\beta]=0$.  Interesting cases depend on the particular physical setting and context. For instance, if the short wave amplitude $a$, as well as the coupling constants $\alpha$ and $\beta$ are not vanishing, the plane wave solution is stable if the wave number $q$ is so chosen to take the value $q=(\beta/2\alpha)-\alpha b$. It is plain that other choices are possible to satisfy the stability condition $r=0$.

 The algebraic construction of the eigenfunctions of the linearised long wave-short wave equation (\ref{linearYON}), according to the method that we have displayed in this work, requires not only the computation of a solution of the YON system (\ref{YON}), but also the computation of the corresponding solution $\Psi(x,t,\lambda)$ of the Lax pair associated to that solution. Generically this task may turn out to be quite difficult; however, it may be feasible for particular solitary wave solutions, either periodic (breathers) or localised solitons.

Few integrable models have been systematically investigated so far by the present method \cite{degasperis2018,degasperis2019,degasperis2019II,romano2021}.
Subsequent research should be devoted to investigate linear stability of solutions of other integrable wave equations of applicative relevance.
In particular, this linear stability approach needs to be formulated so as to deal with solutions which are different from just continuous waves, whereas results obtained in the scalar cases should be extended to multicomponent systems. Our present results are likely to be relevant to the investigation of solutions which are homoclinic to continuous waves, such as solitons propagating on a background, and rogue waves.


\bibliographystyle{ieeetr}

\bibliography{main}

\begin{thebibliography}{10}

\bibitem{dodd1982}
R.~K. Dodd, H.~C. Morris, J.~C. Eilbeck, and J.~D. Gibbon, {\em Solitons and
  Nonlinear Wave Equations}.
\newblock New York: Academic Press, 1982.

\bibitem{degasperis2009}
A.~Degasperis, ``Multiscale expansion and integrability of dispersive wave
  equations,'' in {\em Integrability} (A.~Mikhailov, ed.), vol.~767 of {\em
  Lecture Notes in Physics}, pp.~215--244, Berlin: Springer, 2009.

\bibitem{yajima1976}
N.~Yajima and M.~Oikawa, ``Formation and interaction of sonic-{Langmuir}
  solitons: {Inverse} scattering method,'' {\em Prog. Theor. Phys.}, vol.~56,
  pp.~1719--1739, 1976.

\bibitem{benney1977}
D.~J. Benney, ``A general theory for interactions between short and long
  waves,'' {\em Stud. Appl. Math.}, vol.~56, pp.~81--94, 1977.

\bibitem{newell1978}
A.~C. Newell, ``Long waves-short waves; a solvable model,'' {\em SIAM J. Appl.
  Math.}, vol.~35, pp.~650--664, 1978.

\bibitem{ling2011}
L.~Ling and Q.~P. Liu, ``A long waves-short waves model: {Darboux}
  transformation and soliton solutions,'' {\em J. Math. Phys.}, vol.~52,
  p.~053513, 2011.

\bibitem{calogero2000}
F.~Calogero, A.~Degasperis, and J.~Xiaoda, ``Nonlinear {Schr\"{o}dinger}-type
  equations from multiscale reduction of {PDEs}. {I}. {Systematic}
  derivation,'' {\em J. Math. Phys.}, vol.~41, pp.~6399--6443, 2000.

\bibitem{calogero2001}
F.~Calogero, A.~Degasperis, and J.~Xiaoda, ``Nonlinear {Schr\"{o}dinger}-type
  equations from multiscale reduction of {PDEs}. {II}. {Necessary} conditions
  of integrability for real {PDEs},'' {\em J. Math. Phys.}, vol.~42,
  pp.~2635--2652, 2001.

\bibitem{geng2020}
R.~Li and X.~Geng, ``On a vector long wave-short wave-type model,'' {\em Stud.
  Appl. Math.}, vol.~144, pp.~164--184, 2020.

\bibitem{geng2020a}
R.~Li and X.~Geng, ``A matrix {Yajima} {Oikawa} long-wave-short-wave resonance
  equation, {Darboux} transformations and rogue wave solutions,'' {\em Commun.
  Nonlinear Sci.}, vol.~90, p.~105408, 2020.

\bibitem{geng2019}
X.~Geng and R.~Li, ``On a vector modified {Yajima} {Oikawa} long-wave
  short-wave equation,'' {\em Mathematics}, vol.~7, p.~958, 2019.

\bibitem{ma1978}
Y.~C. Ma, ``The complete solution of the long-wave short-wave resonance
  equations,'' {\em Stud. Appl. Math.}, vol.~59, pp.~201--221, 1978.

\bibitem{baronio2015}
F.~Baronio, S.~Chen, P.~Grelu, S.~Wabnitz, and M.~Conforti, ``Baseband
  modulation instability as the origin of rogue waves,'' {\em Phys. Rev. A},
  vol.~91, p.~033804, 2015.

\bibitem{chen2018}
J.~Chen, Y.~Chen, B.~F. Feng, K.~Maruno, and Y.~Ohta, ``General high-order
  rogue waves of the (1+1)-dimensional {Yajima}-{Oikawa} system,'' {\em J.
  Phys. Soc. Jpn.}, vol.~87, p.~094007, 2018.

\bibitem{chen2019}
J.~Chen, L.~Chen, B.~F. Feng, and K.~Maruno, ``High-order rogue waves of a
  long-wave-short-wave model of {Newell} type,'' {\em Phys. Rev. E}, vol.~100,
  p.~052216, 2019.

\bibitem{degasperis2018}
A.~Degasperis, S.~Lombardo, and M.~Sommacal, ``Integrability and linear
  stability of nonlinear waves,'' {\em J. Nonlinear Sci.}, vol.~28,
  pp.~1251--1291, 2018.

\bibitem{talanov1965}
V.~Talanov, ``Self focusing of wave beams in nonlinear media,'' {\em JETP
  Lett.}, vol.~2, p.~138, 1965.

\bibitem{benjamin1967}
T.~Benjamin and J.~Feir, ``The disintegration of wave trains on deep water.
  {Part} 1. {Theory},'' {\em J. Fluid. Mech.}, vol.~27, pp.~417--430, 1967.

\bibitem{pelinovsky2021}
D.~E. Pelinovsky, ``Instability of double-periodic waves in the nonlinear
  {Schr\"{o}dinger} equation,'' {\em Front. Phys.}, vol.~9, p.~6, 2021.

\bibitem{wright2010}
O.~C. Wright, ``On a homoclinic manifold of a coupled long-wave-short-wave
  system,'' {\em Commun. Nonlinear Sci.}, vol.~15, no.~8, pp.~2066--2072, 2010.

\bibitem{degasperis2019}
A.~Degasperis, S.~Lombardo, and M.~Sommacal, ``Coupled nonlinear
  {Schr\"odinger} equations: {Spectra} and instabilities of plane waves,'' in
  {\em Nonlinear Systems and Their Remarkable Mathematical Structures}
  (N.~Euler and M.~C. Nucci, eds.), vol.~2, pp.~206--248, Boca Raton: CRC
  Press, 2019.

\bibitem{borluk2008}
H.~Borluk and S.~Erbay, ``Stability of solitary waves for three-coupled long
  wave-short wave interaction equations,'' {\em IMA J. Appl. Math.}, vol.~76,
  pp.~582--598, 2008.

\bibitem{erbay2012}
H.~Erbay and S.~Erbay, ``Transverse linear instability of solitary waves for
  coupled long-wave short-wave interaction equations,'' {\em Appl. Math.
  Lett.}, vol.~25, pp.~2402--2406, 2012.

\bibitem{baronio2014}
F.~Baronio, M.~Conforti, A.~Degasperis, S.~Lombardo, M.~Onorato, and
  S.~Wabnitz, ``Vector rogue waves and baseband modulation instability in the
  defocusing regime,'' {\em Phys. Rev. Lett.}, vol.~113, 2014.

\bibitem{degasperis2016}
A.~Degasperis and S.~Lombardo, ``Integrability in action: {Solitons},
  instability and rogue waves,'' in {\em Rogue and Shock Waves in Nonlinear
  Dispersive Media} (M.~Onorato, S.~Resitori, and F.~Baronio, eds.),
  pp.~23--53, Berlin: Springer, 2016.

\bibitem{gardner1968}
R.~M. Miura, C.~S. Gardner, and M.~D. Kruskal, ``Korteweg-{de Vries} equation
  and generalizations. {II}. {Existence} of conservation laws and constants of
  motion,'' {\em J. Math. Phys.}, vol.~9, pp.~1204--1209, 1968.

\bibitem{calogero1982}
F.~Calogero and A.~Degasperis, {\em Spectral Transform and Solitons: {Tools} to
  Solve and Investigate Nonlinear Evolution Equations}, vol.~1.
\newblock Amsterdam: North-Holland, 1982.

\bibitem{degasperis2019II}
A.~Degasperis, S.~Lombardo, and M.~Sommacal, ``Rogue wave type solutions and
  spectra of coupled nonlinear {Schr\"odinger} equations,'' {\em Fluids},
  vol.~4, no.~1, p.~57, 2019.

\bibitem{romano2021}
S.~Lombardo, M.~Sommacal, and M.~Romano, ``The 3-wave resonant interaction
  model: {Spectra} and instabilities of plane waves,'' {\em In preparation},
  2021.

\end{thebibliography}

\end{document}